\documentclass[11pt,twocolumn]{article}
\usepackage[
  top=2cm,
  bottom=2cm,
  left=2cm,
  right=2cm
]{geometry}

\usepackage{comment}

\usepackage[hidelinks]{hyperref}
\usepackage{xcolor}
\usepackage[normalem]{ulem}  % for underlines

\usepackage{markdown}

\definecolor{darkblue}{RGB}{0,0,139}

\renewcommand{\href}[2]{\textcolor{darkblue}{\uline{#2}}}
\usepackage{graphicx} 
\usepackage{makecell}
\usepackage{multirow}
\usepackage{array}
\usepackage{url}
\usepackage[table]{xcolor}
\usepackage{acronym}
\usepackage{booktabs}
\usepackage{longtable} % JF input
\usepackage{tabularx} % JF input 
\usepackage{ltablex} % JF input
\usepackage{caption} % JF input
\usepackage[T1]{fontenc}
\usepackage{float}
\usepackage{amsmath}
\usepackage{booktabs}
\usepackage{array}
\usepackage{ragged2e}
\usepackage{placeins}  % put in preamble

\usepackage{courier}  % or \usepackage{beramono}
\usepackage{booktabs, array, ragged2e}

\usepackage{xspace}

\usepackage{enumitem}
\setlist[itemize]{leftmargin=*, nosep} 

\usepackage[backend=biber,style=ieee,sorting=none,maxcitenames=1,minnames=1]{biblatex}
\title{Cross-National Evidence of Disproportionate Media Visibility for the Radical Right in the 2024 European Elections}

\author{
Íris Damião$^{1,2\dag}$ \and
João Franco$^{1\dag}$\footnote{Current address: Vermont Complex Systems Institute, Burlington, Vermont, USA} \and
Mariana Silva$^{1}$ \and
Paulo Almeida$^{1}$ \and
Pedro C. Magalhães$^{3}$ \and
Joana Gonçalves-Sá $^{1,4}$}
\date{}

\begin{document}
\maketitle

\small
\noindent $^{1}$ Social Physics and Complexity Lab - SPAC, LIP - Laboratório de Instrumentação e Física Experimental de Partículas, Lisbon, Portugal \\
$^{2}$ Instituto Superior Técnico - Universidade de Lisboa, Lisbon, Portugal \\
$^{3}$ Instituto de Ciências Sociais da Universidade de Lisboa, Lisbon, Portugal \\
$^{4}$ NOVA LINCS - NOVA Laboratory for Computer Science and Informatics, Department of Computer Science, NOVA School of Science and Technology, NOVA
University Lisbon, Lisbon, Portugal \\
\\
\noindent $^{\dag}$These authors contributed equally to this work. \\
\\
Corresponding author: Joana Gonçalves-Sá (e-mail: joanagsa@lip.pt) \\

\begin{abstract}
This study provides a systematic comparative analysis of media visibility of different political families during the 2024 European Parliament elections. We analyzed close to 21,500 unique news from leading national outlets in Austria, Germany, Ireland, Poland, and Portugal -- countries with diverse political contexts and levels of media trust. Combining computational and human classification, we identified parties, political leaders, and groups from the article's URLs and titles, and clustered them according to European Parliament political families and broad political leanings. 
Cross-country comparison shows that the Mainstream and the Radical Right were mentioned more often than the other political groups. Moreover, the Radical Right received disproportionate attention relative to electoral results (from 2019 or 2024) and electoral projections, particularly in Austria, Germany, and Ireland. This imbalance increased in the final weeks of the campaign, when media influence on undecided voters is greatest. Outlet-level analysis shows that coverage of right-leaning entities dominated across news sources, especially those generating the highest traffic, suggesting a structural rather than outlet-specific pattern. 
Media visibility is a central resource, and this systematic mapping of online coverage highlights how traditional media can contribute to structural asymmetries in democratic competition.
\end{abstract}

\section{Introduction}

In modern democracies, multiple factors shape electoral outcomes, including party ideology and leadership. Another crucial element is the attention that the media allocates to different parties \cite{arkolu2023, VanRemoortere2023}. For citizens to consider a party as a viable option, they must not only be aware of its existence but also encounter it with sufficient frequency to form at least a minimal understanding of its positions on key issues~\cite{Johann2017}. Media \textit{salience}, or the distribution of media resources (e.g, screen time, newspaper coverage), then contributes to voters’ perceptions of a party’s viability and importance~\cite{Jerit_Barabas_Bolsen_2006, Vliegenthart2012}. Indeed, it has been shown that media salience can ``tip the balance'' among competing parties, with visibility of a political actor being crucial~\cite{Geers2016}.

Early research focused on the attention given to certain issues, positing that the more attention an issue receives, the higher the public perception of its importance \cite{Gerber2009}. Later work refined this concept, framing salience as a multidimensional construct, encompassing \textit{attention} (frequency of mentions), \textit{prominence} (placement and presentation of coverage), and \textit{valence} (the tone of the coverage, typically classified as positive or negative) \cite{Kiousis2004}. Yet, the literature remains mixed on which of these dimensions most strongly shapes electoral outcomes.

Regarding attention, \textcite{Eberl2015} found that frequency biases in Austrian newspaper coverage during the 2013 elections had no immediate effect on voter intention, while noting that they might influence long-term perceptions of party legitimacy. By contrast, a study in Denmark \cite{Hopmann2010} showed that even small variations in exposure can affect voter intentions, especially among the undecided. 
%In the United States, local media visibility was shown to enhance voter knowledge, political moderation, and turnout \cite{snyder2010press,MYERS2025}.

Regarding valence, a positive tone is generally considered a key factor in vote switching \cite{Geers2016}. However, \textcite{VanRemoortere2023} argue that positive coverage has little effect, whereas negative coverage harms electoral performance, while other scholars, such as \textcite{Vliegenthart2012}, find that visibility alone matters for electoral success. Moreover, analyzing the interactions between valence and visibility is even more complex, as the effects of tonal direction often prove mixed or counter-intuitive. Studying German elections, \textcite{Gei2017} showed that while tone improved the explanatory power of voting models, excessive positive coverage could actually reduce support among decided voters. Conversely, and if frequent, negative coverage may boost attention and engagement, producing a ``no bad publicity'' effect \cite{deVreese2006_2, Mancosu2021}, increasing support for the criticized actors. 

%In Spain, research spanning many years of  publications, from the two main newspapers in the country, found that the media systematically prioritizes leading parties and the mainstream opposition over smaller parties, with governing parties receiving more coverage but also more negative tone, reinforcing existing power structures through attention patterns \cite{Baumgartner2015}. 

%Even the type of media platform can also affect political visibility and, consequently, its effects on the electoral results. Finally, platforms matter: television tends to amplify visibility for established parties, while newspapers often provide entry points for lesser-known actors \cite{Gei2017,Vermeer2022}.

These interactions are further compounded by the rise of social media and online platforms, which have altered both the production and consumption of political information. We highlight three key effects. First, through their social media pages, political actors, parties, and citizens can elevate issues, some of which are later picked up by traditional media \cite{Gilardi2021}, and the agenda-setting becomes multi-directional \cite{Chadwick2017, chadwick2015politics}, with political communication unfolding within hybrid media systems. Second, trust in the media has eroded substantially~\cite{gallup_trust_media_2025,statista308468}. Political actors, particularly populist leaders, may exploit this by emphasizing perceived asymmetries in visibility and accusing mainstream outlets of censorship or ideological bias~\cite{Krmer2018, Figenschou2018}. Such claims resonate with voters who perceive a mismatch between their own concerns and media priorities, creating a ``salience gap'' that fuels distrust and populist identification ~\cite{Vliegenthart2012, Fawzi2018}. Third, social media platforms have been increasingly used to bypass the same traditional media accused of gatekeeping, allowing fringe or underrepresented parties to gain visibility~\cite{schmidt2020party}. These platforms have been criticized for fostering echo chambers and ideological homophily, reinforcing polarization and amplifying extremist views through engagement-driven algorithms~\cite{ChuecaDelCerro2024},~\cite{LevyRazin2022},~\cite{Arora2022}. Importantly, since online news has become the dominant source of information~\cite{reuters2025digitalnews}, traditional media have also become subject to similar engagement pressures.
Together, hybrid media systems, declining trust, and persistent gatekeeping claims contribute to a broader agenda shift toward topics favored by radical parties, which extend to mainstream media coverage~\cite{SaldiviaGonzatti2025, ellinas2007playing}.

Reflecting this trend, the European political landscape provides a particularly salient context to examine such dynamics. There is growing evidence that visibility of party leaders and salience of issues such as immigration in the traditional media across Europe, lead to increased support for anti-immigration parties \cite{Dennison2019, Boomgaarden2007}. Research conducted in Belgium, Germany, and the Netherlands~\cite{Vliegenthart2012} further shows that such parties gained both visibility and electoral success when the media increased coverage of issues such as migration and cultural identity -- issues these parties are commonly perceived to “own.” Notably, the tone of this coverage does not need to be positive to contribute to normalization: \textcite{Murphy2018} finds that even predominantly negative media coverage benefited the UK Independence Party, which at the time was a fringe actor within a first-past-the-post electoral system.

% And, in the UK, \textcite{Murphy2018} found that even negative coverage benefited the UK Independence Party, which was, at the time, a fringe party in a first-past-the-post system.

Beyond national politics, comparative research shows that media salience can extend across borders. Coverage of foreign radical right actors affects perceptions of their domestic prevalence and shapes normative beliefs about the movement more broadly~\cite{selvanathan2025farright}.Related work on transnational diffusion demonstrates that domestic political attitudes can polarize following foreign party success, with national media acting as a key transmission channel~\cite{bohmelt2024antiimmigration}. Finally, ~\textcite{vandewardt2024contagion} show how ideological families spread across borders through emulation (party family contagion), often well before organizational entry or electoral consolidation takes place. Therefore, media-driven agenda shifts may propagate transnationally, reinforcing support for radical right parties even in countries where they were previously marginal.

% This also suggests that traditional media continues to play a decisive role in shaping political competition in multiparty systems. How they cover populist actors can either constrain or amplify populist influence, with even negative or critical reporting potentially contributing to the normalization of fringe movements by granting them sustained visibility \cite{stier2024mainstreaming, udris2012populist}: attention may outweigh negative valence, particularly for undecided voters and rising extremist parties. 
In summary, visibility alone -- irrespective of tone -- is politically meaningful and appears to be a powerful driver of political relevance, at least for some political groups.

% But how does the visibility allocated to different political actors vary across European media? 

Given this context, this study examines how media visibility of political leanings varies across countries in the lead-up to the 2024 European Parliament election, offering a comparative perspective.

We analyzed the frequency of mentions of political actors and parties in the 2024 EU election across five countries: Austria, Germany, Ireland, Poland, and Portugal. These countries were chosen for their differing early-2024 polling trends and varying trajectories of the Radical Right. In the 2019 European Parliament election, the Radical Right had minimal presence in Portugal and Ireland, was rapidly growing in Austria and Germany, and was well-established in Poland. At the time of the study, none of the five countries had Radical Right or Radical Left parties in government.

We focus on online news media for two reasons:(1) Traditional media outlets increasingly engage audiences through their online platforms~\cite{Chadwick2017} with online news largely reflecting the key topics of their broadcast and print content; and (2) despite being less trusted \cite{eurobarometer2022media}, they represent the most common source of news consumption in all five countries -- ranging from 66\% in Germany to 84\% in Poland, according to the Reuters Digital News Report~\cite{reuters2025digitalnews}-- surpassing television, print, and radio in every case. 

\begin{figure*}[ht!]
    \centering
    \includegraphics[width=1\textwidth]{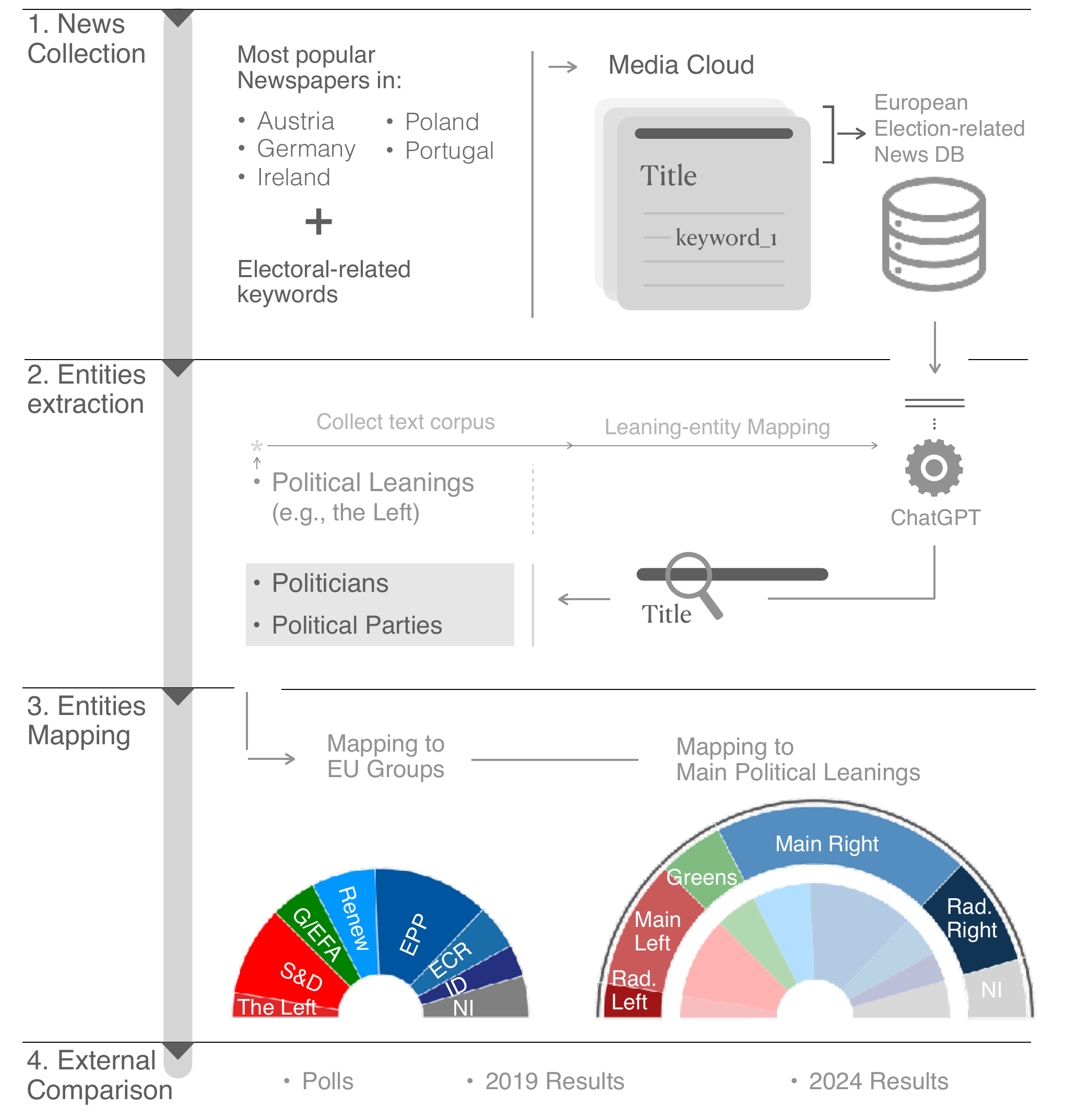}
    \caption{\textbf{Methodological Pipeline.} News extraction and classification was done in four sequential steps. (1) Following electoral-related keyword searches on Media Cloud, news articles were collected. (2) Through automatic classification, identified political entities (politicians, parties, political leanings) were matched to their respective national parties. Only news with mentions in their titles or URLs were included. (3) National entities were matched to their European Parliament groups (as defined in 2019) and then to five main political leanings (from left to right: Radical Left - dark red, Mainstream Left - light red, Greens - green, Mainstream Right - light blue, and Radical Right - dark blue). The new far-right formations emerging in 2024  -- Patriots for Europe and Sovereignists -- were classified under the Identity and Democracy group for comparability. (4) Frequency of media mentions were compared to the 2019 and 2024 electoral results and to the 2024 pooled voting intentions.}
    \label{fig:methodology}
\end{figure*}

We used Media Cloud to collect articles related to the European Parliament elections during the two months preceding the vote, a period in which most voters typically make their electoral decisions, many only in the final weeks \cite{le2019vote}. From these articles titles and URLs, we extracted all relevant political entities (see Figure~\ref{fig:methodology}). We measured media visibility at the level of party families within each country by counting mentions not only of national parties but also of other European members of each party family. This option reflects our view of political competition in Europe today, in which information about political actors abroad can still matter domestically, even when no clear organizational or electoral counterparts exist at home. We then compared the proportion of media attention allocated to each political group with previous election results, polling estimates, and post-election outcomes.

Our findings indicate that the Radical Right received disproportionately high media attention relative to both its prior representation and its expected or eventual electoral performance. Notably, the Radical Right emerged as the most frequently mentioned segment of the political spectrum even in countries such as Ireland, where it has no parliamentary representation.

Taken together, these results suggest the presence of a common pattern in media coverage across national contexts, which may be contributing to the normalization and legitimization of extremist movements in Europe.

\section{Methodology}

A comprehensive news database was built to assess media attention toward political actors across the European Parliament’s ideological spectrum in five countries: Austria, Ireland, Germany, Poland, and Portugal. Media frequencies were compared with broader political trends (electoral results and predictions) -- Figure \ref{fig:methodology}.

\subsection{Media outlet pre-selection}

Media sources were chosen based on online visibility, per country, according to the free version of \textit{Semrush Traffic Analytics}. This is a software-as-a-service platform for online salience and content marketing [\href{https://www.semrush.com}{\textit{Semrush}}], that ranks website visits in specific months, geographies, and categories. Ranks were collected in February 2025, reflecting the most visited media sources during January 2025. The selected category was ``Newspapers'', with the top 20 results collected for each country (e.g., \href{https://www.semrush.com/website/top/austria/news-and-media/}{\textit{Semrush - Top 20 Newspapers in Austria}}). As these also included broadcast media, social media, and search engine pages, such as TV channels, \texttt{Instagram}, and \texttt{Google}, only traditional media and dedicated online news websites were kept (\ref{fig:methodology_news_collection}). %We kept the broadcasters but removed the social media / search engine entries, restricting the list to traditional or digital news media.  %and restricted our lists to websites from media sources with presence in traditional media (newspapers, television channels, radio). 

\subsection{News collection and database processing}

\begin{figure*}[!ht]
    \centering
    \includegraphics[width=1\textwidth]{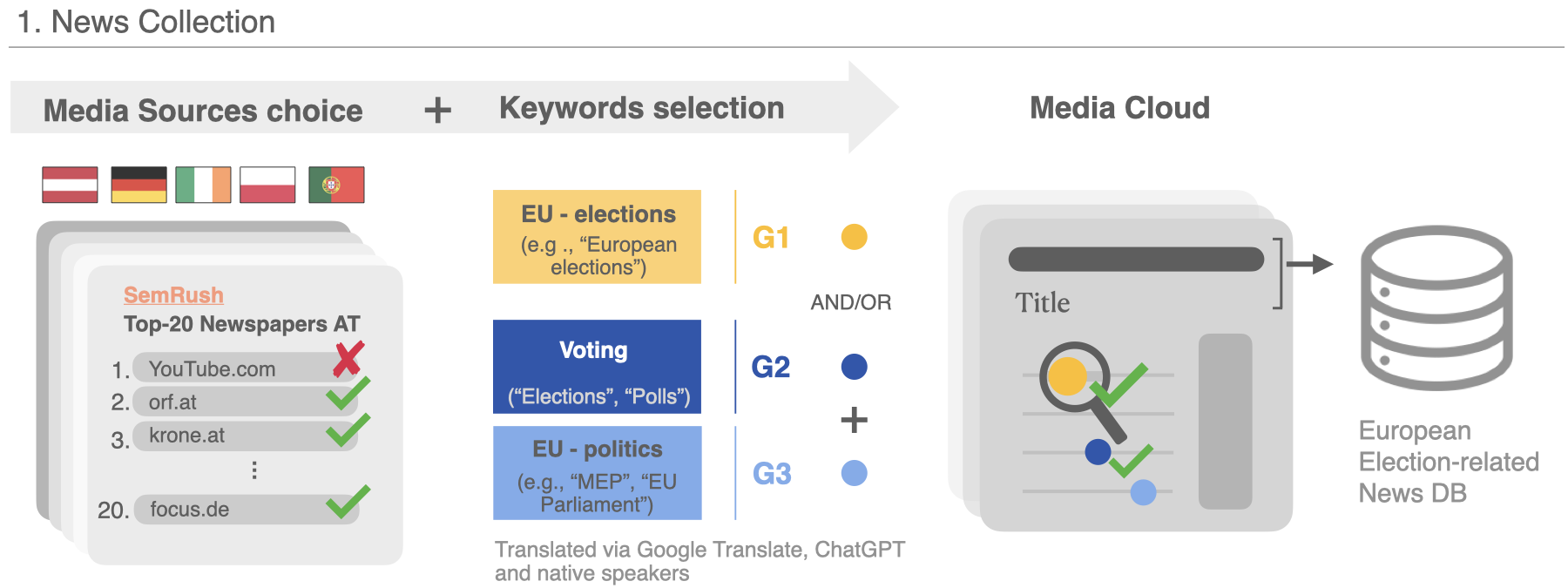}
    \caption{\textbf{Methodological Pipeline - News Collection.} Media Sources were selected from \textit{Semrush} Top 20 Newspapers for each country and excluding social media and search engine pages (e.g., ``YouTube.com''). News articles were collected from these sources using two sets of queries (yellow and blue), derived from three lists of terms in English, German, Polish, and Portuguese. The titles and URLs of all news articles from the domains identified through the \textit{Semrush} lists, and containing at least one query from each set, were gathered to form the final Europe-related News Database.
    Videos and news in other languages, returned by \textit{Media Cloud}, were filtered out and not included.}
    \label{fig:methodology_news_collection}
\end{figure*}

News were collected using [\href{https://www.mediacloud.org/}{\textit{Media Cloud}}], an open-source platform that aggregates web content from global media sources, through the [\href{https://www.mediacloud.org/media-cloud-search}{\textit{Media Cloud Search tool}}] and its API (\ref{fig:methodology_news_collection}), during March 2025 and concluded on March 28\footnote{As detailed in Section \ref{section:eu_elections_related_news_database}, news related with the second round of Polish local elections had to be collected. These were only collected on May 1st 2025 based on related Polish queries and media outlets.}.   
Extraction was limited to the previously identified media sources, in the two months preceding the election, from April 9 to June 9, 2024 (or June 7, in the case of Ireland). Keywords were defined as combinations of terms, with case sensitivity ignored. 
Queries can be done based on media sources, time period, and keywords present in the webpages' content. Although searches covered the entire webpage, including title, lead, main text, and sidebars, \textit{Media Cloud} only returns the article title and respective URL to comply with copyright restrictions. 

%\medskip

\subsubsection{Outlet processing and final selection}
%\medskip
Two steps were taken to ensure consistency in outlet selection: domain and location analysis.

As searches were performed on the top-ranking domains identified by \textit{Semrush}, multiple outlets under the same domain could be included in the \textit{Media Cloud} results. In some cases, these additional sources corresponded to regional editions or specific sections of the queried outlet (e.g., the business section), and were included in the analysis. 
% In Portugal, this issue also affected distinct media sources with similar domains: for example, querying \href{https://sapo.pt/}{sapo.pt} (which was identified as popular \textit{Semrush}), also returned results from \href{https://eco.sapo.pt/}{eco.sapo.pt} (an independent outlet, on the list). In this case, the latter news were eliminated.

%\medskip
Location was a factor in two different aspects. First, \textit{Semrush} identified popular media outlets that were not registered in the country at hand (e.g., UK-based outlets in Ireland). Second, in other instances, \textit{Media Cloud} returned location-specific domains, such as \href{https://www.theguardian.com/uk}{The Guardian UK} and \href{https://www.theguardian.com/us}{The Guardian US}. Decisions were made on a case-by-case basis. 
Popular news outlets from different countries were included when they used the same main language (UK outlets in Ireland, German in Austria, Brazilian in Portugal). In the case of the same outlets appearing in two different countries (e.g., \textit{spiegel.de} in Germany and Austria), the collected news were included in both countries. When \textit{Media Cloud} returned both geography-specific editorial domains and the parent domain (e.g., \href{https://g1.globo.com/ba/bahia/}{g1.globo.com/ba/bahia/} and \href{https://www.globo.com/}{globo.com}), or considered specific URLs as individual media sources, only the parent domains were selected. In the case of multiple locations of the same domain, the geographically closer was chosen (\href{https://www.theguardian.com/uk}{The Guardian UK} for Ireland). 
Finally, some media sources were identified by \textit{Semrush} and are mapped in \textit{Media Cloud}'s list of sources, but the keyword searches did not return any election-related news (Table \ref{tab:media_outlets_media_cloud_nonews}). These are not mentioned in the database.

%\medskip

Table \ref{tab:tab_media_sources_considered} lists the media sources identified by \textit{Semrush} (gray), with the news collected from \textit{Media Cloud} (bold), along with their cumulative share of the country's overall visits. For each country, these cover a broad range, from public broadcasters to "elite" newspapers and tabloids.
%Media sources (and respective news articles) associated with more than one country (e.g., The Guardian) were associated with all relevant countries.

\begin{table*}[ht!]
\centering
\caption{Top 20 media outlets in \textit{Semrush} rankings for each country, excluding social media platforms and search engines. Media sources shown in \textbf{bold} are those for which news articles were retrieved from Media Cloud. Column \textbf{N sources w/ News} reports the total number of outlets per country with news retrieved. Column  \textbf{Share of total visits} indicates the percentage of total visits these outlets represent relative to the \textbf{"Top \textit{Semrush}"} list.}
\label{tab:tab_media_sources_considered}

\renewcommand{\arraystretch}{1.2} % increases row height for readability
\setlength{\tabcolsep}{6pt} % adjusts horizontal padding
\small
\begin{tabular}{
    l
    >{\RaggedRight\arraybackslash}p{9cm}
    >{\RaggedRight\arraybackslash}p{2cm}
    >{\RaggedRight\arraybackslash}p{2.8cm}
}
\toprule
\textbf{Country} & \textbf{Top \textit{Semrush}} & \textbf{N sources w/ News} & \textbf{Share of total visits} \\
\midrule
Austria &
{\ttfamily\bfseries bild.de}, {\ttfamily\bfseries derstandard.at}, {\ttfamily\bfseries diepresse.com}, \texttt{exxpress.at}, {\ttfamily\bfseries focus.de}, \texttt{heute.at}, \texttt{kleinezeitung.at}, {\ttfamily\bfseries krone.at}, {\ttfamily\bfseries kurier.at}, \texttt{meinbezirk.at}, {\ttfamily\bfseries n-tv.de}, {\ttfamily\bfseries nachrichten.at}, {\ttfamily\bfseries orf.at}, {\ttfamily\bfseries spiegel.de}, {\ttfamily\bfseries tt.com}, {\ttfamily\bfseries vol.at}
& 12 & 85.3\% \\
\addlinespace
Germany &
{\ttfamily\bfseries bild.de}, 
\texttt{finanzen.net}, \texttt{fr.de}, {\ttfamily\bfseries faz.net}, {\ttfamily\bfseries focus.de}, {\ttfamily\bfseries mdr.de}, {\ttfamily\bfseries merkur.de}, {\ttfamily\bfseries n-tv.de}, {\ttfamily\bfseries spiegel.de}, \texttt{sueddeutsche.de}, {\ttfamily\bfseries tagesschau.de}, {\ttfamily\bfseries tagesspiegel.de}, {\ttfamily\bfseries t-online.de}, {\ttfamily\bfseries welt.de}, {\ttfamily\bfseries zeit.de}
& 12 & 92.5\% \\
\addlinespace
Ireland &
{\ttfamily\bfseries bbc.co.uk}, {\ttfamily\bfseries bbc.com}, {\ttfamily\bfseries dailymail.co.uk}, \texttt{globo.com}, {\ttfamily\bfseries independent.ie}, {\ttfamily\bfseries irishexaminer.com}, {\ttfamily\bfseries irishtimes.com}, {\ttfamily\bfseries nytimes.com}, {\ttfamily\bfseries rte.ie}, \texttt{rip.ie}, {\ttfamily\bfseries sky.com}, {\ttfamily\bfseries telegraph.co.uk}, \texttt{thesun.ie}, {\ttfamily\bfseries theguardian.com}, \texttt{tvn24.pl}
& 11 & 90.4\% \\
\addlinespace
Poland &
{\ttfamily\bfseries dorzeczy.pl},  \texttt{fakt.pl}, {\ttfamily\bfseries gazeta.pl}, \texttt{kwejk.pl}, \texttt{niezalezna.pl}, \texttt{o2.pl}, \texttt{plejada.pl}, \texttt{pomponik.pl}, \texttt{pudelek.pl}, {\ttfamily\bfseries polsatnews.pl}, \texttt{se.pl}, {\ttfamily\bfseries sport.pl}, {\ttfamily\bfseries rmf24.pl}, {\ttfamily\bfseries tvn24.pl}, {\ttfamily\bfseries wpolityce.pl}, {\ttfamily\bfseries wyborcza.pl}
& 8 & 50.1\% \\
\addlinespace
Portugal &
\texttt{abola.pt}, {\ttfamily\bfseries cmjornal.pt}, {\ttfamily\bfseries expresso.pt}, \texttt{flashscore.pt}, {\ttfamily\bfseries globo.com}, {\ttfamily\bfseries iol.pt}, \texttt{jn.pt}, {\ttfamily\bfseries noticiasaominuto.com}, \texttt{ojogo.pt}, {\ttfamily\bfseries observador.pt}, {\ttfamily\bfseries publico.pt}, {\ttfamily\bfseries record.pt}, {\ttfamily\bfseries rtp.pt}, {\ttfamily\bfseries sapo.pt}, \texttt{tempo.pt}, \texttt{zerozero.pt}
& 10 & 67.6\% \\
\bottomrule
\end{tabular}
\end{table*}

\subsubsection{News extraction}

Identifying \ac{EU} election-related articles relied on two keyword-based heuristics. First, three groups of words were created.
%Our focus was salience within the scope of \ac{EU} elections. Hence, we had to collect collect all news articles that discuss these elections. Different approaches have been used to identify discussed topics in text [INSERT]. Considering the limitations in accessing the overall news content (title, lead and main text) through \textit{Media Cloud}, we defined an article as approaching \ac{EU} elections based on two simple heuristics.
%\medskip
%Second, articles may engage with \ac{EU} elections more subtly -- for example, through non-contiguous mentions of election-related and EU-related terms (e.g., ``Europe is going to vote''). Other combinations, such as election terms with candidate names or with the election date, may also indicate \ac{EU} elections-related articles. However, these cases likely overlap with the combination just described above and were excluded to maintain a concise keyword list. 
%\medskip
Group 1 (EU-elections) includes words that explicitly mention the \ac{EU} elections at least once (e.g., ``European elections'' or ``European Parliament elections'').
Group 2 (Voting) includes broad election-related terms (e.g., ``elections'', ``polls,'' ``campaign'').
Group 3 (EU-politics) includes European Parliament-related terms (e.g., ``European Parliament'', ``Members of the European Parliament'', ``MEP'').

%\medskip
As querying only from Group 1 risks false negatives (as the elections might be mentioned without using these specific n-grams), but querying articles using terms from Groups 2 and 3 alone risks false positives, as election-related terms may appear in non-political contexts (e.g., ``sports club elections''), and references to the European Parliament may concern non-electoral activities (e.g., ``a speech delivered in the European Parliament''), a combinatorial approach was used. 
To be considered, news needed to include at least one keyword from Group 1 -- explicitly linked to the European elections -- or one keyword from Group 2 and another from Group 3. 
%non-contiguous search of all pairwise combinations of terms from the general election-related and European Parliament-related groups. 

%\medskip

These keywords were defined in English and translated into each country's official language, except in Ireland's case, using online tools (Google Translate and ChatGPT), and reviewed by native speakers -- ranging from one to five individuals per language. These assessed linguistic accuracy and contextual relevance to electoral and election-related language and were also encouraged to suggest additions or removals, accounting for language-specific morphological features, including declensions (gender, nominative, and others).
%For example, Portuguese terms reflect gender through declension. In Polish, words undergo declensions depending on grammatical case -- nominative (mianownik: rzeka), genitive (dopełniacz: rzeki), dative (celownik: rzece), accusative (biernik: rzekę), instrumental (narzędnik: rzeką), locative (miejscownik: rzece), and vocative (wołacz: rzeko). Other declensions are used in German. Queries keywords needed to account for all possible declensions
Keywords were searched-for individually, to avoid possible false positives from ``fuzzy'' matches and character wildcards, which would be difficult to identify from the item's title and URL. Table \ref{sup_table: mediacloud_queries} shows the final list of terms, divided into the three groups and by language, used to query \ac{EU} related articles. 

\subsubsection{EU elections-related news database}\label{section:eu_elections_related_news_database}

\textit{Media Cloud} returns news articles that contain the provided keywords, including matches in the title, main body, or sidebar references to other news. Results from webpages categorized as “video” or “videos,” whether in English or in the country’s official language, were removed as the keywords were typically identified in the sidebar and not in the text of the main news piece.

%Webpages categorised as ``videos'' were removed because their respective keywords were manually identified as frequently occurring in sidebar content, related to other news items. 

%From these, only articles with entities mentioned in the titles or URLs were included in the analysis (approximately half, see below and Figure~\ref{fig:methodology_entities_extraction}).

%Results of webpages categorised as ``video'' or ``videos'', in English or in the official language of the country, were removed, since keywords were manually identified to be frequently present on the sidebar, as \textit{Media Cloud} cannot access keywords mentioned within the video itself. 

News articles in languages other than English, German, Polish, and Portuguese (as reported by \textit{Media Cloud}) were also removed, corresponding to 0.2\% of the total collected news (Table~\ref{tab:news_entities_country}).
%an indicator for the robustness of our approach.

%\medskip

Finally, and during the considered 2-month period, Irish local elections and the second round of Polish local elections also took place (on June 7th and April 21st, respectively). To remove the salience impact resulting from these elections, an exclusion mechanism was implemented: articles containing exclusion keywords were removed from the set of articles retrieved by the ``target keywords,'' unless they were returned by queries explicitly linked to the European elections. Table~\ref{sup_table:mediacloud_queries_exclusion} lists all keywords used in the exclusion mechanism. %No other political campaigns were identified in the target countries. 

The final database includes 21,528 news items that were then analyzed for the presence of political entities.

%\medskip

% In the end, 22,578 news articles were retrieved (Table \ref{tab:tab_stats_news}).

 % \begin{table}[h!]
 % \centering
 % \caption{Number of collected news articles}\label{tab:tab_stats_news}
 % \begin{tabular}{|c|c|} \hline
 % \textbf{Country} & \textbf{\# News articles} \\ \hline
 % Austria  & 5,642 \\ \hline
 % Germany  & 3,861 \\ \hline
 % Ireland  & 3,779 \\ \hline
 % Poland   & 3,372 \\ \hline
 % Portugal & 6,441 \\ \hline
 % Total    & 23,095 \\ \hline
 % \end{tabular}
 % \end{table}

\subsection{Political entities identification}\label{entities_extraction}

To identify relevant political entities in the headlines and URLs of the compiled news database, two complementary approaches were employed: (a) extraction using a Large Language Model (LLM), and (b) fuzzy-matching, followed by (c) manual validation. 
For the LLM approach, entities included candidates and elected representatives from the 2019 and 2024 elections (as listed by \href{https://www.europarl.europa.eu/meps/en/full-list}{Europarl}), national and European political parties, and general references to the political spectrum (e.g., left, center, right, extreme). For fuzzy matching, the process was the same except that only elected representatives and respective parties during the 2024 elections were included.
Searches included the full names of parties and political-spectrum terms in both English and the respective official languages (e.g., Socialist Party and Partido Socialista), while individuals were matched using the most frequently used names, typically first and last (obtained from \href{https://www.europarl.europa.eu/meps/en/full-list}{Europarl}).

\begin{figure*}[!ht]
    \centering
    \includegraphics[width=1\textwidth]{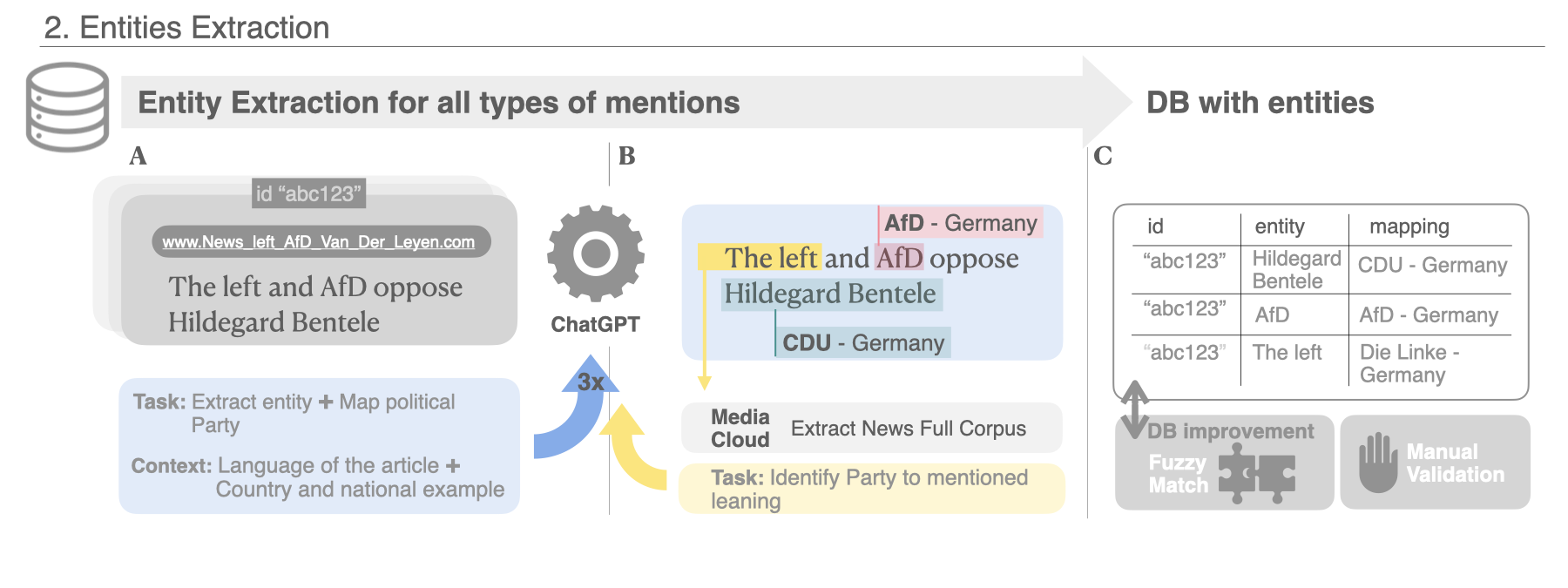}
    \caption{\textbf{Methodological Pipeline - Entities Extraction. A)} All news headlines and their respective URLs were processed using the ChatGPT-4o API to identify political entities -- such as politicians (e.g., Hildergard Bentele), political parties (e.g., AfD), and political leanings (e.g., the left). \textbf{B)} For mentions of political leanings (yellow), the news corpus was further searched to determine the specific political entity referred to under that label. \textbf{C)} Fuzzy matching was applied to detect entities missed by ChatGPT, and a manual validation of 150 news items was conducted to assess extraction accuracy.}
    \label{fig:methodology_entities_extraction}
\end{figure*}

\subsubsection{LLM-based classification}

ChatGPT-4o was instructed to 1) extract references to political entities: politicians, political parties, political groups, or political leanings, from either the URL, the headline, or both; and 2) to associate identified entities with specific political parties and countries when applicable (e.g: politicians $\rightarrow$ party \textbf{+} country; party $\rightarrow$ country). The prompt used for extraction was adapted to each national context, accounting for (1) the expected language of the news articles, and (2) illustrative examples provided to guide the model’s output. Each country prompt was run three times (using OpenAI API default temperature), and political entities were included if returned at least once, to reduce false negatives caused by variability in the outputs of the \ac{LLM}. The exact prompts can be found in the Supplementary Information (Table \ref{sup_table:prompt_political_entities_extraction}).

%\medskip

By prompting ChatGPT-4o to extract political entities, these were listed as they appear in the article. For example, references to politicians could use only first, only last, both first and last, or two last names, while political parties could be mentioned as acronyms, or full or partial names. These extracted entities were manually evaluated and standardized (e.g, ``Von der Leyen'' and ``Ursula von der Leyen'' as references to ``Ursula Von der Leyen'', ''Müller", as possibly referring to either ''Alexander Müller" or ''Piotr Müller", depending on context). 
When the identified mentions corresponded to areas of the political spectrum (e.g., ``right''), an additional step was taken. Specifically, the full article content was retrieved using the Python package \texttt{\textit{Media Cloud} Metadata Extractor}~\cite{mediacloud-metadata}. Then, the retrieved text, together with the URL and headline, were resubmitted to ChatGPT-4o with instructions to identify the specific political parties or politicians referenced (if any), using only the information provided in the article (strict prompt in Figure~\ref{sup_fig:identify_party}, Appendix).

After having all types of entities associated with a national political party (e.g. ``Von der Leyen'' to ``CDU''), these associations were (1) manually validated (using \textit{Wikipedia} and the EP official website \href{https://www.europarl.europa.eu/meps/en/full-list}{Europarl}), and (2) standardized to the English version of all parties names (e.g., ``CDU - Germany'' to "Christian Democratic Union - Germany'' or ``Vox - Spain'' to ``Voice - Spain''), as reported in Table~\ref{tab:national_parties_leaning} of the Appendix. 

In most cases, the association between party mentions and national party identification was straightforward, with few cases of possible ambiguity. In particular, titles/URLs where the same name/acronym could refer to two or more possible leanings (e.g.: ``CDU'' as Germany's ``Christian Democratic Union'', aligned with the EPP-European People's Party, or Portugal's ``Unitary Democratic Coalition", aligned with The Left in the European Parliament). If no other mentions were made (e.g., to politicians) and only the acronym was used, political entities were allocated according to the country of the news item (i.e., `CDU'' as Germany's ``Christian Democratic Union'' if the article was published in Germany).

\subsubsection{Fuzzy matching}\label{section:fuzzy_matching}

The fuzzy matching algorithm was implemented with the function \texttt{fuzz.partial\_ratio} from the \texttt{fuzzywuzzy} Python library. 
The function leverages the Ratcliff/Obershelp pattern recognition algorithm, implemented via Python’s \texttt{SequenceMatcher} to compute partial string similarity. For each pair of input strings, the function determines the shorter ($s_{1}$) and longer ($s_{2}$) ones (e.g., ``Socialist party'', ``Socialism'') and identifies all common substrings (e.g, ``Socialis''), recording the starting indices in both strings. Using this alignment, the function extracts a substring ($K_{m}$) from the longer string that matches the length of the shorter one ($s_{1}$), starting from the aligned index (e.g, ``Socialist''), and identifies the number of matching characters ($M_{m}$) between $K_{m}$ and the shorter string ($s_{1}$). Finally, it calculates a similarity score between the shorter string and this extracted segment  \({\displaystyle D_{s1,s2} = 2 \cdot M_{m} \div (|s_{1}| + |s_{2}|)}\), $|s_{1}|, |s_{2}|$ as the length of $s_{1}, s_{2}$. This process is repeated for all matching blocks and the highest resulting similarity score is returned, scaled to a 0–100 range. 

%\medskip

Both provided strings and sub-strings (e.g.: ``Socialist Party'' -> ``Socialist Party'' + ``Socialist'' + ``Party'') were considered during the fuzzy matching approach. Matches with fuzzy score equal or above 90 for individual name components (e.g., “Socialist” and “Party”) and above 80 for full names (e.g., “Socialist Party”) were manually reviewed. The entities with other scores were filtered out.

%\medskip

During manual validation, additional political entities were found and also incorporated and their names standardized as before. This was consistent with the goal of fuzzy matching, which aimed to maximize the detection of relevant entities rather than to assess the original performance against a fixed input list.

%\medskip

In total, only $\sim$5\% of the items were identified by this approach alone (494 unique new items), meaning that 94.2\% of the entities were equally identified by both (fuzzy matching and ChatGPT-4o), suggesting classification robustness.

%As there was no clear bias in misidentification, this performance was deemed adequate and fuzzy matching was not extended to other political entities or other parts of the articles.

\subsubsection{Performance assessment}

Performance of the semi-automated entity extraction (ChatGPT-4o \& Fuzzy matching) was assessed, based on manual validation of 150 randomly selected news articles by 10 individuals. Coders were instructed to confirm (or not) the accuracy of the identified entities and to report possible omissions. The number of articles reviewed by each individual varied according to their language proficiency. 

Overall, the evaluation yielded an overall accuracy of $\approx 95\%$. The weakest performance (75\% accuracy) was obtained when classifying political leaning (left, center, or right), before applying the additional step of retrieving article text excerpts and resubmitting them to ChatGPT-4o for explicit party identification. After this additional step, a separate validation of 100 items was carried out and showed a near-perfect performance, except for mentions to ``center'' and ``extremist''. These specific cases were manually reviewed and matched to the correct political party. In roughly 20\% of the leaning-related mentions, no political party could be confidently identified, and these items were excluded from the analysis.
% If no party could be confidently assigned, the item was not included. 
No misclassification bias regarding specific languages, party groups, or politicians was identified.

\subsection{Political leaning classification}

The previous steps resulted in a list of political parties and European political groups. All national parties were subsequently associated with the correspondent European family (Figure \ref{fig:methodology}, ``Entities Matching''), using Wikipedia information, the official EU Parliament official website [\href{https://www.europarl.europa.eu/about-parliament/en/organisation-and-rules/organisation/political-groups}{link}], and the official websites of all European Groups (e.g., \href{https://www.eppgroup.eu/}{EPP official website}).
%Politicians were first associated with their national parties, at the time of the elections, based on Wikipedia information (Figure \ref{fig:methodology_entities_extraction}, right table). 

% \subsubsection{Party identification and selection}
%\subsubsection{Party filtering}
\subsubsection{Party processing and final list}

The list of entities and associated political parties obtained in the previous step (section~\ref{entities_extraction}) was diverse and included parties not relevant to this study (e.g., ``Labour Party - UK'', ``Republican Party - USA''), as they were not participating in the 2024 European elections. Therefore, the final list of political parties was manually filtered using the following non-mutually exclusive criteria:

\begin{enumerate}
    \item Mentions of political parties. These had to be associated with EU Member States at the time (excluding third party countries and the UK) and (a) be running in the 2024 EU Parliament Election, or (b) having been elected in 2019. 
    \item Mentions of political groups. These had to be represented in the European Parliament during the 2019-2024 term~\cite{eu_political_groups}.
\end{enumerate}

% Note that, in most cases, the association between party mentions and national party identification was straightforward. Two types of situations required special handling: (1) entity ambiguity, typically if the same name refers to two or more possible leanings (e.g.: ``CDU'' as Germany's ``Christian Democratic Union'' - Mainstream Right -, or Portugal's ``Coligação Democrática Unitária'' (Unitary Democratic Coalition) - Radical Left). In the few instances where only the acronym was used or if it was not clear to which entity the news were referring to, simple heuristics were implemented, allocating political entities according to the country of the news item. 
In some cases, entities that satisfied the criteria mentioned above appeared in articles from unrelated countries (e.g.: ``AfD'', a political party from Germany, being mentioned in Irish or Polish news). In this case, the articles were included in the analysis with the leaning corresponding to the leaning of the entity. This means that, for example, mentions to Radical Right parties in Irish newspapers were counted as Radical Right even if they mention parties or politicians from other locations (see Table~\ref{tab:national_parties_leaning} in Appendix). 
%one of the other four countries but not if they mention entities from other locations).  

The resulting list included mentions to 150 different national parties, from 26 EU countries. All parties in the five countries at hand that have elected \ac{MEPs} in either the 2024 or the 2019 elections are represented. 
%. These ambiguous cases were rare, allowing for targeted adjustments and corrections. On other occasions, 
%\medskip

%ChatGPT-4o failures did not correlate with specific entities. ``Sebastião Bugalho'' and ``Lena Schilling'', the most frequent entities failed to identify by ChatGPT-4o, corresponded to 1 and 0.5\% of failures respectively, and to 28\% and 29\% of reported ``Sebastião Bugalho'' and ``Lena Schilling''.

%\medskip

\subsubsection{Leaning mapping}

The national political parties were mapped to European parliamentary groups and then the European parties were grouped into five broad political leanings.

First, a correspondence was found between each national party and one of the established European Parliament groups from the 2019 to 2024 legislative term: The Left; S\&D – Progressive Alliance of Socialists and Democrats; G/EFA – Greens/European Free Alliance; Renew – Renew Europe Group; EPP – European People’s Party (Christian Democrats); ECR – European Conservatives and Reformists; ID – Identity and Democracy; and NI – Non-attached (Figure 1, panel 3 and Supplementary Table~\ref{tab:national_parties_leaning}). New entities, or entities linked to newly formed alliances (not present in the previous 2019-2024 Parliament configuration) were evaluated manually. Namely, the Patriots for Europe and Europe of Sovereign Nations were classified under the Identity and Democracy (ID) group, as both derive from, and align ideologically with, the former Identity and Democracy parliamentary group \cite{FarRightReconfiguration}.

In parallel, each national party was classified by political ideology using the 2024 Chapel Hill Expert Survey~\cite{ches2024}, which draws on assessments from 69 political scientists to position 279 parties across 31 countries, on dimensions including ideology, populism, democracy, EU integration, and specific policy stances. Based on this data, parties were categorized into 10 ideological groups: Radical Right, Conservative, Liberal, Christian-Democratic, Socialist, Radical Left, Green, Regionalist, Confessional, and Agrarian/Center.
Therefore, each national party was classified both in terms of political ideology (according to the Chapel Hill survey) and as belonging to a EU parliamentary group during the 2019-2024 term.

%With guidance from a political scientist collaborator, 

Finally, each European Parliament party was assigned to one of five broader ideological groups: \textbf{Radical Left}, represented in dark red, \textbf{Mainstream Left} (Socialist), in light red, \textbf{Greens}, in green, \textbf{Mainstream Right} (including Conservative, Liberal, and Christian-Democratic), in light blue, and \textbf{Radical Right}, in dark blue in all figures. 

As these groups may include national parties classified as belonging to different families, the mode was used to assign a party classification. For example, if one group includes one party classified as Regionalist, another as Nationalist, and three more as Radical Right, this group would be assigned to the Radical Right (EU parliament mapping scheme in Figure~\ref{fig:methodology}, full party matching in Supplementary Table~\ref{tab:national_parties_leaning}).

%This approach enabled the classification of both explicit references to European political families and of parties whose ideological alignment was not immediately evident: the latter were indirectly classified based on their affiliation with a European Parliament group.

\subsection{Analysis}
\subsubsection{Mention counts and temporal dynamics}\label{analysis_mentions}

All extracted mentions were assigned to one of the five broader political leanings. Unique mentions of each group were counted per news item and aggregated at the country level. These counts were used to compute the proportion of mentions associated with each political leaning in each country: 

\begin{equation}
P_{c,\ell} = \frac{M_{c,\ell}}{M_{c,\text{total}}} \times 100
\end{equation}

Where $M_{c,\ell}$ denotes the number of mentions associated with political leaning $l$ in country $c$, and $M_{c,\text{total}}$ represents the total number of political mentions extracted for that country. 

At the individual country level, analysis includes both a temporal evolution of mentions (Figure ~\ref{fig:figure_over_time}) and the full 2-month aggregate (Figures~\ref{fig:figure_results} and ~\ref{fig:figure_leaning_all}).

To obtain an overall measure across all countries, a weighted average was calculated based on each country's share of seats (i.e., their representation in the European Parliament), for the five countries included in the analysis. Let $S_c$ be the number of seats allocated to country $c$, and $S_{total} = \sum_c S_c$. The weighted average proportion for each political leaning $l$ is therefore: 

\begin{equation}
\overline{P}_{\ell} = \sum{c} \left( \frac{S_c}{S_{\text{total}}} \times P_{c,\ell} \right)
\end{equation}

\subsection{External sources comparison}

Mention frequency was analyzed in reference to three external criteria: (a) 2019 European Parliament election results \cite{EuropeanParliament2019ConstitutiveSession}, (b) pre-election polling data from \href{https://euobserver.com/eu-elections/areaba2f2f}{EUobserver}~\cite{euobserver2025euelections}, published during the week preceding the election day, and (c) official 2024 electoral results\cite{EuropeanParliament2024Results}. 

All measures were computed individually for each country, with national political parties grouped using the previously described procedure (see section~\ref{analysis_mentions}).

Differences between the proportions of mentions and each of the three reference values were computed as: 
\begin{equation}
    Diff = P_{c,\ell} - E_{c,\ell}
\end{equation}
where $E_{c,\ell}$ represents one of the three possible reference (``External'') values: 2019 results, poll data, or 2024 results. 

Difference $Diff$ magnitudes were classified as being less than one, less than two, or more than two standard deviations away from the mean value of $P_{c,\ell}$ (shown in Figure~\ref{fig:figure_results} as dark gray shading, light gray shading, or no shading, respectively).

\subsection{Outlet-level analysis}

News outlets were classified in four different dimensions:

\begin{enumerate}
    \item \textit{Dominant Leaning:} for each outlet, the \textit{dominant political leaning} was determined as the category (Radical Left, Greens, etc.) occurring most frequently among all news items collected from that outlet (Figure~\ref{fig:figure_popularity}, circle color). 
    
    \item \textit{Leaning Bias:} indicates the proportion of the outlet’s total news items that was classified as being in the dominant category. Three categories were created from 20\% (close to equal distribution among the five categories) to more than 75\% (almost all mentions were to the dominant leaning)-- Figure~\ref{fig:figure_popularity}, circle size.

    \item \textit{Popularity:} Popularity was operationalized using each outlet’s ranking in the country-specific \textit{Top 20} \textit{Semrush} traffic ranking. After excluding social media platforms and search engines, the mean rank position across outlets was 8. Outlets ranked above 8 were considered \textit{high-popularity}, while those ranked 8 or lower were considered \textit{lower-popularity} -- Figure~\ref{fig:figure_popularity}, x-axis.

    \item \textit{Output production:} Production volume was assessed by computing the mean number of news items published per outlet across all media sources. Outlets publishing fewer news items than the mean were classified as \textit{low-output outlets}, whereas outlets producing more than the mean were classified as \textit{high-output outlets} -- Figure~\ref{fig:figure_popularity}, y-axis.
\end{enumerate}

The last two classifications created four quadrants of combinations of output and popularity (Table ~\ref{tab:quadrant_learnings}). For each of these quadrants, the percentage of outlets in which each political leaning was the most frequently mentioned, in headlines and URLs, was calculated.

\section{Results}
\subsection{Media coverage of the EP Elections was broad and increased with time}

To study the visibility that European media give to different political parties we first identified popular news websites in each of the five countries -- Austria, Ireland, Germany, Poland, and Portugal --, according to a digital marketing platform (\textit{Semrush}, see Methods for details). This selection included different types of outlets, such as television channels, newspapers, radio stations, and specialized broadcasters (e.g., finance or sports), provided they produce news and maintain dedicated websites for their publication. 
Table~\ref{tab:tab_media_sources_considered} shows the breadth of sources (at least 8 per country).

\begin{figure}[!ht]
    \centering
    \includegraphics[width=1\columnwidth]{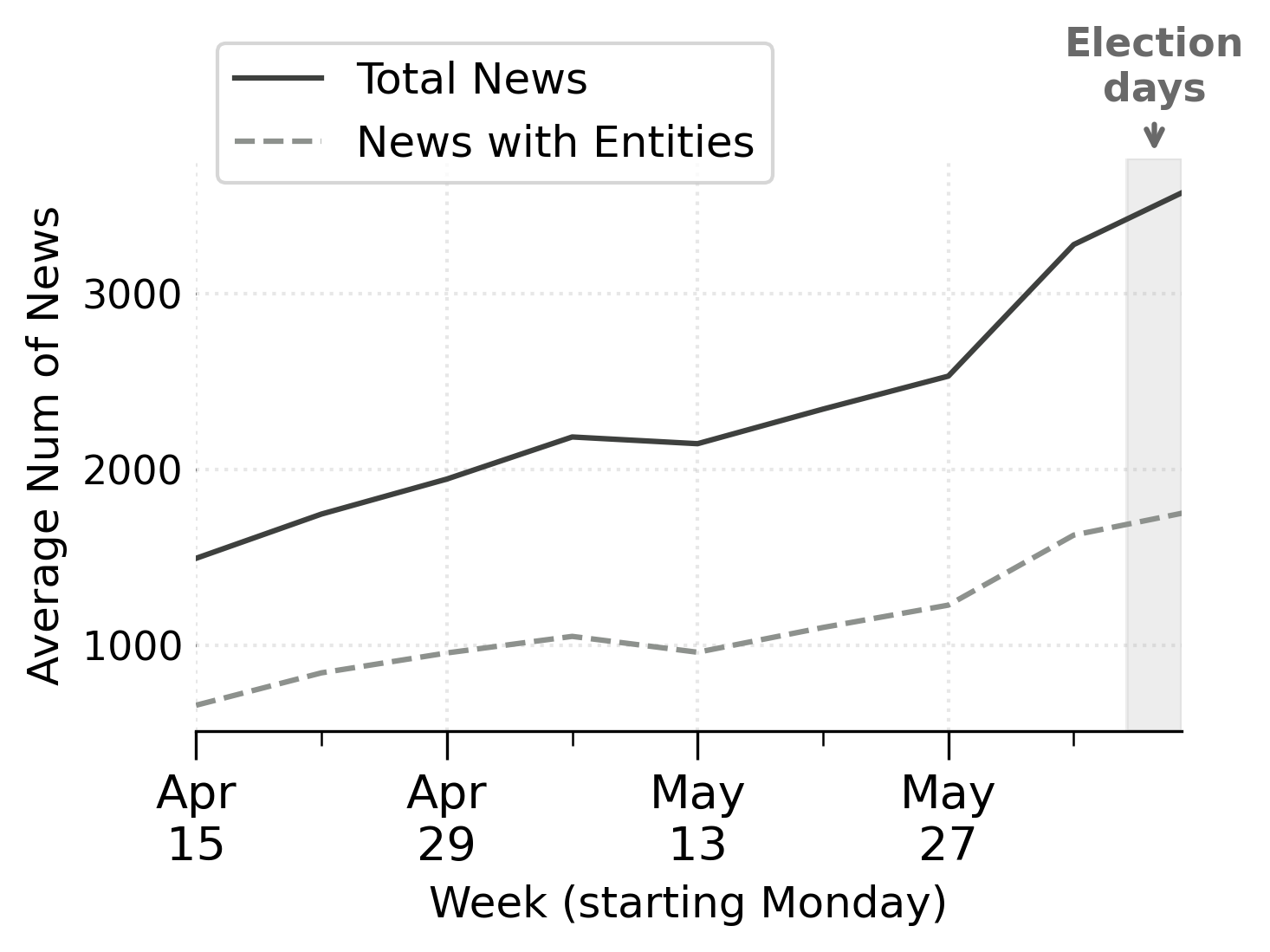}
    \caption{Average weekly number of news items from \textit{Media Cloud} (solid line) and the subset with mapped entities (dashed line). Vertical lines mark two-week intervals. A 3-week running window was used to compute the averages.}
    \label{fig:news_over_time}
\end{figure}

We then used \textit{Media Cloud}, an open-source platform that aggregates web content from global media sources, and applied a combinatorial system of keywords to identify all news items referring to the European Parliament elections, published by these outlets (see Methods and Supplementary Table~\ref{sup_table: mediacloud_queries} for details).  Across the five countries, the elections received consistent coverage, with a steady increase throughout the analysis period (black line in Figure~\ref{fig:news_over_time}). 
%Some comment on temporal trends?
%some comment of which had more mentions? I know we do not have a baseline to say if they represet x% of the total number of news.
In total, we extracted 21,528 unique news items, with the distribution per country varying between approximately 3,300 in Poland and 6,400 in Portugal (Table~\ref{tab:news_entities_country}). 

\begin{table}[ht!]
\centering
\caption{News items per country. Total number of extracted news, number of unique news items with political entities in the titles / URLs and percentage of news items that had political entities in the titles /URLs. Five-country totals in last row.}
\begin{tabular}{c c c}
\hline
\textbf{Country} & \makecell[l]{\textbf{Total}\\\textbf{News}} & \makecell[l]{\textbf{Unique News} \\\textbf{with entities (\%)}}\\
\hline
Austria  & 3,480 & 1,868 (53.7\%) \\
Germany  & 3,861 & 2,156 (55.8\%) \\
Ireland  & 4,449 &   593 (13.3\%) \\
Poland   & 3,297 & 2,234 (67.8\%) \\
Portugal & 6,441 & 3,441 (53.4\%) \\
\hline
\textbf{Total} & 21,528 & 10,292 (47.8\%) \\
\hline
\end{tabular}
\label{tab:news_entities_country}
\end{table}

% It is important to note that some of the collected news items may correspond to false negatives, as a direct consequence of how \textit{Media Cloud} operates. The platform retrieves articles whose HTML content includes the queried keywords. Consequently, a keyword may appear in the HTML not because it is part of the article itself, but because it is embedded elsewhere on the webpage (for example, in links or previews of other stories). As a result, some articles identified by \textit{Media Cloud} might not actually discuss the elections. To control for this we...

\subsection{Mainstream and Radical Right Entities receive more visibility}

\begin{figure*}[!ht]
    \centering
    \includegraphics[width=0.93\textwidth]{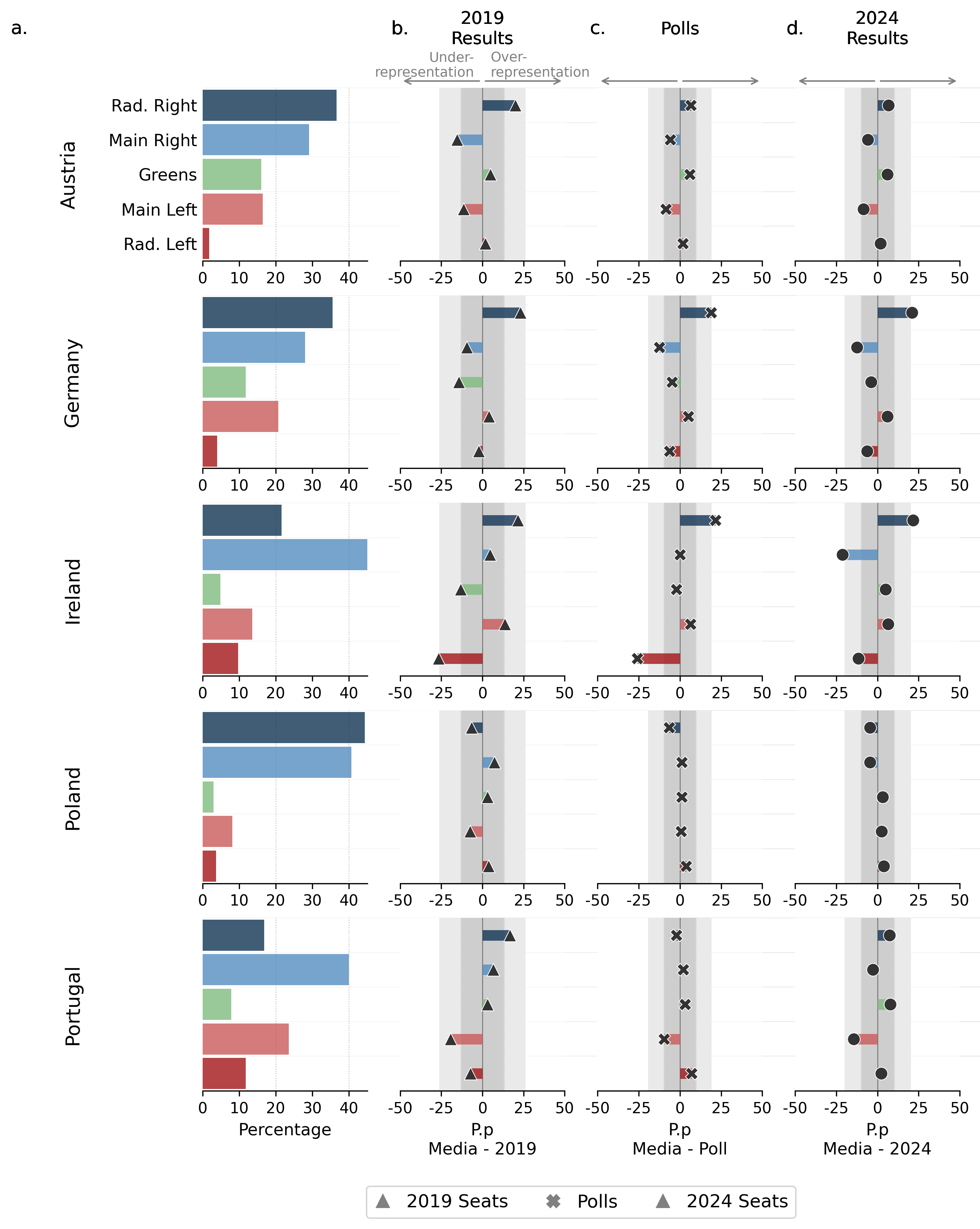}
    \caption{\textbf{a)} Share of political entity mentions in the news, grouped by major European political families (as defined in Scheme \ref{fig:methodology}), for each country. \textbf{b)} Difference between media attention (Panel a) and the 2019 European Parliament seat distribution (triangles). \textbf{c)} Difference between media attention and pre-election seat projections from EUobserver polls (crosses)~\cite{euobserver2025euelections}. \textbf{d)} Difference between media attention and the final 2024 seat distribution (circles). Symbols within the light grey area fall within one standard deviation of the cross-country mean difference; those in the white area are within two standard deviations.}
    \label{fig:figure_results}
\end{figure*}

Given the large number of identified unique articles, we used a semi-automated, systematic quantification of political entity references. A popular Large Language Model (\ac{LLM}) was provided with the URLs and titles of the extracted news and prompted to identify parties, politicians and political leanings related to the EU Parliament elections. To improve accuracy, articles were grouped and processed by country, using information on existing national entities as training data. Classification was repeated at least three times and fuzzy matching was implemented to complement the \ac{LLM} approach. 94.2\% of entities were equally identified by both methods and a random sample was further subject to human validation (see Methods for details and Figure~\ref{fig:methodology_entities_extraction}). Overall, the estimated accuracy in entity identification was $\approx 95\%$. %Most errors involved mentions of political leaning (left, center and right)
From the more than 21,000 news items that mentioned the EU elections, around 50\% had direct mentions to political entities in their titles or URLs, and these were the ones analyzed (see Table~\ref{tab:news_entities_country}). This conservative approach (focusing on titles/ URLs) serves two main purposes. First, titles are used as a proxy for high visibility, reducing the likelihood of including articles with only superficial mentions, in the news body. Second, since many analyzed outlets are behind paywalls, this method approximates the experience of non-subscribing readers (the vast majority of the population), who often engage only with headlines.

% \begin{figure*}[!ht]
%     \centering
%     \includegraphics[width=1\textwidth]{figures/figure_1.pdf}
%     \caption{\textbf{a} Distribution of political entity mentions across the main political families (as in Scheme~\ref{fig:methodology}), with stacked bars showing the European political families each entity belongs to (based on the 2019 European Parliament composition). \textbf{b} Proportion of entities by main political leaning over time. Values represent weekly averages, using a 3-week rolling window, with each week starting always on a Monday for the 2 months preceding the EU parliament elections. }
%     \label{fig:figure_1}
% \end{figure*}

We then sequentially grouped these classified entities according to 1) their national parties, 2) these parties' European parliamentary group during the 2019-2024 term, 3) the parties political ideology using the 2024 Chapel Hill Expert Survey~\cite{ches2024} and, from 2) and 3), we mapped them into 4) their broad political leaning, divided into Radical Left, Mainstream Left, Center/Greens, Mainstream Right, and Radical Right, as detailed in Figure~\ref{fig:methodology}. Only entities associated to national parties that were running for the 2024 EP elections and had elected at least one Member of the European Parliament (MEP) in 2019 or/and 2024 were considered. A complete list of all political parties identified in the collected news set, along with their corresponding final leaning classification, is provided in Appendix~\ref{national_parties_and_leaning}, Table~\ref{tab:national_parties_leaning}.

Supplementary Table~\ref{tab:top_entities} and Figure~\ref{fig:eu_parties_countries} show the distribution of mentions per national party and European families and the bars in Figure~\ref{fig:figure_results} - \textbf{a}, show this distribution by political leaning, per country (from Austria -- top --, to Portugal -- bottom -- , in alphabetic order). If all families were mentioned similarly, the bars would be equally sized at around the 20\% mark. However, as can be observed, the radical (dark blue) and the Mainstream Right (light blue) received between 57\% (in Portugal) and 85\% (in Poland) of all mentions. Conversely, the Left (light red) and the Radical Left (dark red) never sum more than 35\% (in Portugal). In fact, the Radical Right entities received the majority of mentions in Austria, Germany, and Poland and the second most in Ireland. In total, entities aligned with the Radical Right were mentioned in 31\% of all articles extracted through \textit{Media Cloud}.

\subsection{Visibility cannot be explained by predicted or past results}

We then asked whether this visibility could be explained by the a) presence of these parties in the national governments, b) seat distribution in 2019-2024 term, c) 2024 seat distribution projections, for each country, according to \href{https://euobserver.com/eu-elections/areaba2f2f}{EUobserver}, and d) 2024 results.

At the time of the elections (June 7th in the case of Ireland and June 9th for all other countries), Austria, Portugal, Poland and Ireland had governments led by parties affiliated with the European People’s Party (EPP), a Mainstream Right family. Germany was led by a party affiliated with the Progressive Alliance of Socialists and Democrats (S\&D), classified as Mainstream Left, in coalition with Center/Greens parties. In the case of Poland, the government included a coalition with Mainstream Right and Mainstream Left parties, both identified as moderate. In Ireland, in addition to the Mainstream Right Fianna Fáil, the government was also formed in coalition with a Center/Green party. 
Therefore, the large proportion of mentions to the Mainstream Right could be justified by their high presence in national governments; yet, visibility did not consistently align with government participation. For example, Mainstream Right parties led the governments of both Portugal and Austria, but their comparative visibility was 12 percentage points lower in Austria and comparable to that of the same-family parties in Germany, which were not in government, at the time. Notably, despite receiving a large number of mentions, the Radical Right was not part of the governments of any of these countries during the electoral period.

We then analyzed how the media mentions compared to the other criteria listed above. Figure~\ref{fig:figure_results} right side panels show the deviations from center, with center corresponding to the proportion of media mentions to each party, per country. Shifts to the right mean that the media mentioned that party family (rows) more than what could be expected when comparing it to (columns) the 2019 seats (triangles), the 2024 projections (crosses), and the 2024 results (circles). Shifts to the left mean that the media mentioned those political entities less. The shaded areas mark one (dark) or two (light) standard deviations from the mean. Symbols outside of the shaded areas correspond to a difference of at least two standard deviations. As observed, the Radical Right received more attention than expected from all three criteria in Austria, Germany, and Ireland. In Portugal, it received slightly less than expected when compared to the 2024 projections, as these had a particularly optimistic forecast for this political family. The only exception to this trend was Poland, where the Radical Right is well established but received fewer mentions than expected by all criteria; however, it still had the highest absolute number of mentions among the countries analyzed (1526 times, corresponding to a 0.45 proportion). All other party families received levels of attention that were closer to expectations, except for the Mainstream Left in Austria and in Portugal, which consistently received systematically fewer mentions (symbols shifted to the left). 

Given the five countries’ differing population sizes and corresponding European Parliament representation, the analysis was repeated by aggregating results across all countries, weighting each country’s counts according to its European Parliament seats (larger for Germany and Poland, smaller for the others; see Methods).

\begin{figure}[ht!]
    \centering
    \includegraphics[width=1\columnwidth]{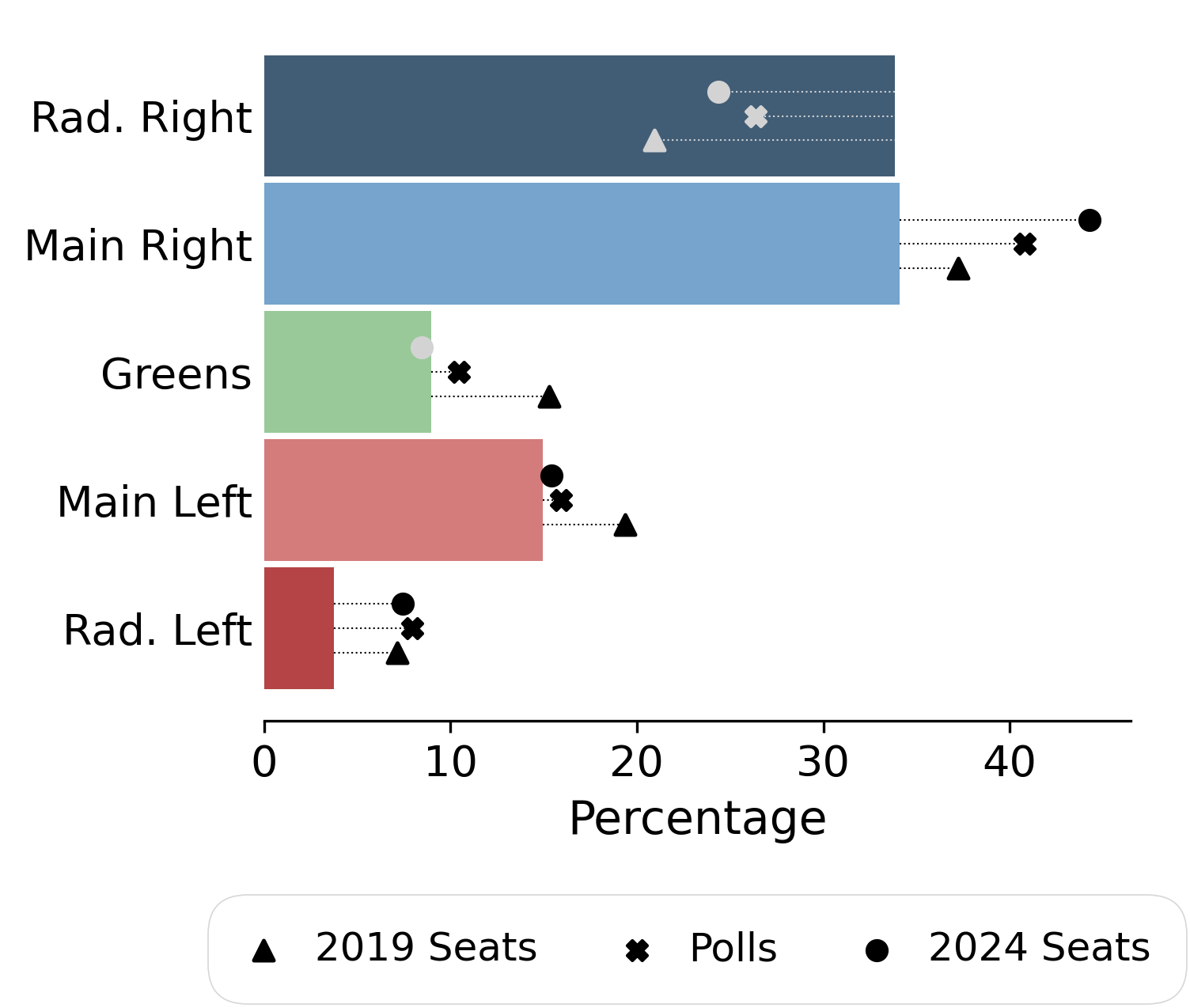}
    \caption{Bars depict the media salience of each political group, weighted by the seat contribution of each country. Triangles correspond to the weighted results from the previous election, crosses denote weighted polling estimates, and dots represent the current weighted distribution of seats across political groups.}
    \label{fig:figure_leaning_all}
\end{figure}

The colored bars in Figure~\ref{fig:figure_leaning_all} show the final weighted results for media salience (i.e. how much attention was given to each political leaning). The overlaid symbols plot the weighted contributions of each country, for the same criteria as before: previous electoral results (triangle), polls (crosses), and 2024 results (circles). When the symbols appear within (to the left of) the bars (in gray), it indicates that the media gave more salience to those groups, whereas symbols outside the bars (or to the right, in black), denote under-representation. The Radical Right remains the most overrepresented, even after accounting for Poland's large seat contribution according to the three criteria (see Figure~\ref{fig:figure_results}). We also note that no other political group showed an overall overrepresentation, even when considering only mentions of national entities (see Supplementary Figure~\ref{fig:eu_parties_countries_weighted}). The Mainstream Right was the most underrepresented, followed by the Greens and the Radical Left. 

\subsection{Radical and Mainstream Right compete for visibility in the weeks leading to the election} 

\begin{figure}[ht!]
    \centering
    \includegraphics[width=0.97\columnwidth]{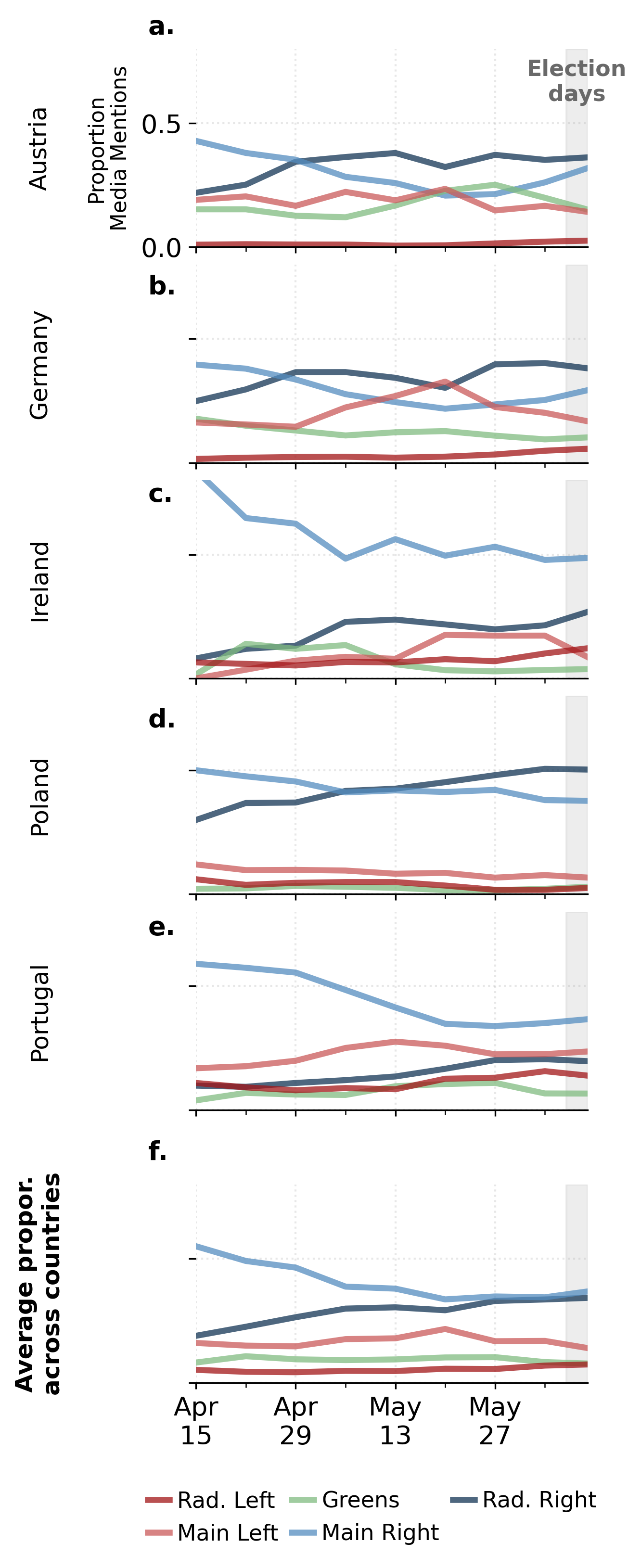}
    \caption{Proportion of entities by main political leaning over time. Values represent weekly averages, using a 3-week rolling window, with each week starting always on a Monday for the 2 months preceding the EU parliament elections. }
    \label{fig:figure_over_time}
\end{figure}

The previous analysis includes all mentions during the two months prior to the EP elections. As it is established that many voters only decide in the few weeks before the election and this is when media salience is expected to have more influence, we analyzed how these mentions varied over time. Figure~\ref{fig:figure_over_time} shows a rolling window for each country (panels \textbf{a} to \textbf{e}, from Austria to Portugal) and the weighted average of all countries (panel \textbf{f}). 

% and as an aggregate of all countries (panel www). there was some varience in time ans across countries but 
%TO BE DONE, see if there is syncronicity

Mentions of Mainstream Right remained high but declined over time, while references to the Radical Right increased in all countries. From May 13 onward, the Radical Right became the most mentioned political leaning in Austria, Germany, and Poland. Mentions to centrist and left-of-center parties were lower and more stable throughout. 

\subsection{The Right dominated across news outlets' popularity}

\begin{figure}[ht!]
    \centering
    \includegraphics[width=1\columnwidth]{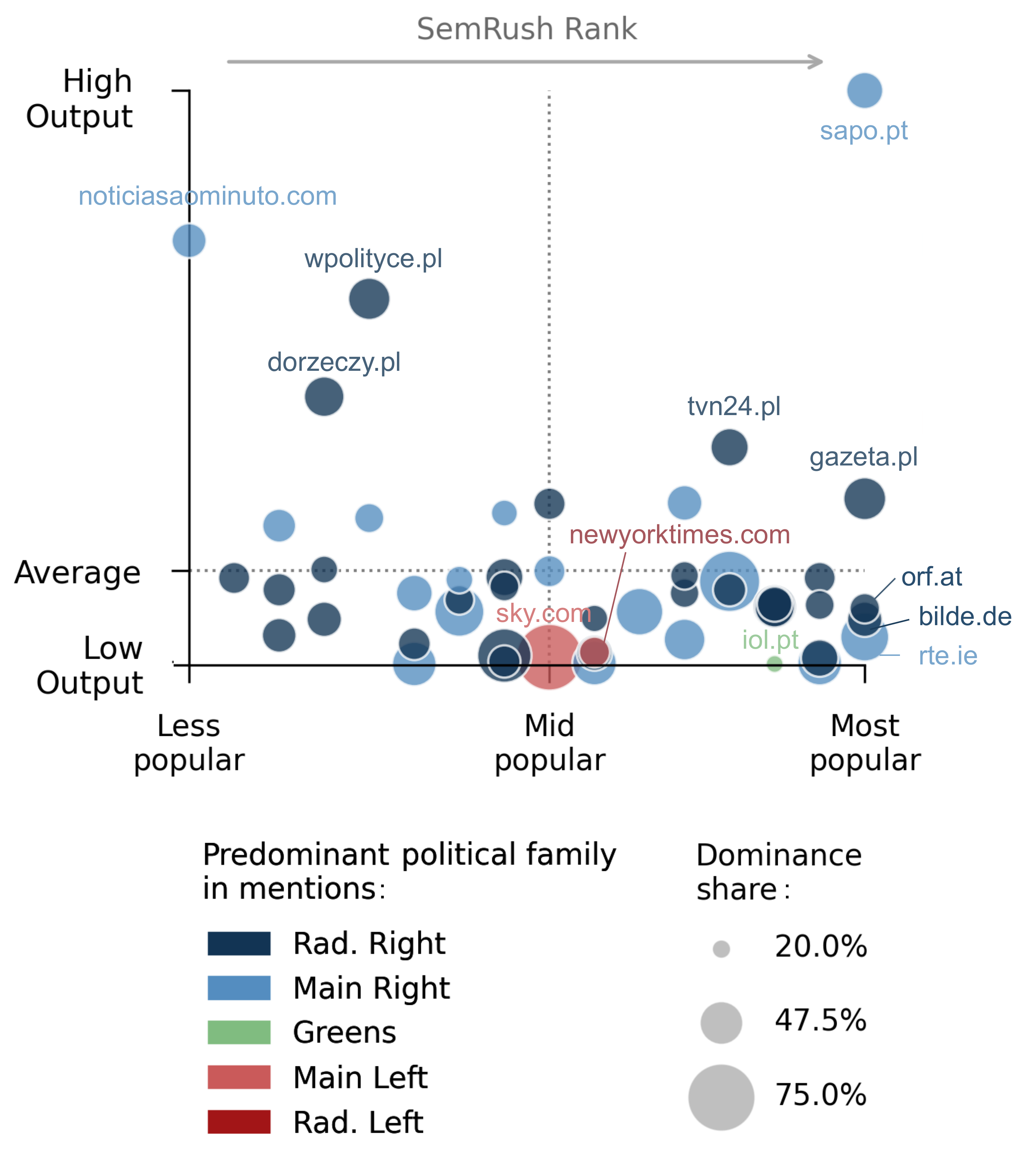}
    \caption{Dominant political leanings of entities mentioned in news headlines across outlets’ popularity and publishing output volume. Each circle represents a media outlet, positioned by its \textit{Semrush} rank (x-axis) and its normalized number of political articles published (y-axis). Colors denote the dominant political orientation of the entities mentioned in that outlet’s headlines, while circle size indicates the share of articles featuring that orientation (i.e., dominance strength).}
    \label{fig:figure_popularity}
\end{figure}

Finally, we examined whether the high salience of the Radical Right -- and of right-leaning actors more broadly -- was concentrated in a few highly productive outlets or reflected a more general media trend. Figure~\ref{fig:figure_popularity} shows, for each outlet (circles), the most mentioned (``dominant'') political family (circle color). As these media sources have very different audience sizes and frequencies of publication, the figure also shows how these ''leanings" vary with outlet popularity (x-axis, according to \textit{Semrush} ranking), and with their publication volume (y-axis, normalized between 0 and 1 relatively to the highest publication volume). Circle size represents the share of articles mentioning the dominant political family: the smallest circles correspond to just over 20\% of all mentions, while the largest indicate that at least 75\% of the outlet’s EU election coverage referenced that family in titles or URLs. Two main patterns emerge: 1) popular outlets generally publish less election articles, though a few very high-traffic outlets publish very frequently, and 2) right-leaning mentions (blue and dark-blue) dominate across both high- and low-popularity media, indicating that the rightward visibility bias extends across the media landscape, not just in fringe outlets. 

Examining Figure~\ref{fig:figure_popularity} by grouping outlets according to popularity and publication volume -- Q1 (bottom-left): less popular, low output; Q2 (bottom-right): more popular, low output; Q3 (upper-left): less popular, high output; Q4 (upper-right): more popular, high output -- we observe a consistent dominance of the Radical Right across the media spectrum (Table~\ref{tab:quadrant_learnings}). In other words, the Radical Right is the dominant leaning in the majority of the outlets, even when analyzed by quadrant: in over 60\% of all news outlets in Q1 and Q2, and 50\% in Q3 and Q4. The Mainstream Right is the dominant leaning in approximately one third of the outlets in Q1 and Q2, tying with the Radical Right in the sparser Q3 and Q4. The Mainstream Left was the dominant leaning in only 6.3\% (N=1) of the outlets in Q1, 5\% (N=1) in Q2, and never in Q3 and Q4. The Greens were the dominant leaning in a single outlet in Q2, while the Radical Left was never the most mentioned leaning in any outlet.

%In Q1, the Radical Right is the dominant leaning in over 60\% of all  in this quadrant, followed by the Mainstream Right (31.3\%) and the Mainstream Left (6.3\%). In Q2, where the majority of outlets cluster, the Radical Right remains dominant with just over 60\% of outlets mentioning this political spectrum most frequently, while the Mainstream Right accounts for roughly one-third of all outlets. Green and Mainstream Left leanings being predominant in under 5\% of outlets. In the higher-output quadrants, the pattern becomes more balanced: in both Q3 and Q4, Radical and Mainstream Right each dominate in 50\% of outlets, and no other leanings appear were mentioned more often by any outlet.
Overall, these results indicate that right-oriented visibility, particularly of the Radical Right, pervades headlines across outlets of all popularity and publication frequency, highlighting a structural asymmetry in the prominence of political actors during EU election coverage.

\begin{table}[h!]
\centering
\caption{Dominant political leanings by outlet popularity and publication volume. Percentage column indicates the proportion of outlets within each quadrant for which a particular political leaning was the most frequently mentioned in headlines and URLs.}
\label{tab:quadrant_learnings}
\small
\begin{tabular}{
    >{\raggedright\arraybackslash}p{2.8cm}  % Quadrant column
    >{\raggedright\arraybackslash}p{3.0cm}  % Leaning column
    >{\raggedright\arraybackslash}p{1.4cm}  % Share column
}
\toprule
\textbf{Quadrant} & \textbf{Dominant Political Leaning} & \textbf{Percentage} \\

\midrule
Q1: Less popular / Low Output & Rad. Right & 62.5 \\
                               & Main Right & 31.3 \\
                               & Main Left & 6.3 \\
\midrule
Q2: More popular / Low Output & Rad. Right & 61.9 \\
                               & Main Right & 28.6 \\
                               & Greens & 4.8 \\
                               & Main Left & 4.8 \\
\midrule
Q3: Less popular / High Output & Main Right & 50.0 \\
                                & Rad. Right & 50.0 \\
\midrule
Q4: More popular / High Output & Main Right & 50.0 \\
                                & Rad. Right & 50.0 \\
\bottomrule
\end{tabular}
\end{table}

\section{Discussion}

This work examined how popular news outlets allocated attention to different political families, in five EU countries -- characterized by distinct political and social contexts -- in the months preceding the 2024 EU Parliamentary elections.
A clear and consistent pattern emerged: Radical Right parties and politicians received a disproportionate share of media attention. Despite substantial differences in national political landscapes, these entities systematically received greater attention than their previous and current electoral performance would anticipate. This imbalance was particularly striking in the case of Ireland, where the Radical Right remains electorally marginal yet appeared in over one third of all analyzed news items. The only exception to this pattern was Poland: there, Radical Right parties do hold a significant share of votes and despite 40\% of all news mentioning these entities -- corresponding to the highest absolute frequency among these countries --, these values were slightly lower than their actual electoral strength. We also observed that Radical Right visibility increased as the election date approached, becoming the most mentioned political leaning, in absolute terms, in Austria, Germany, and Poland. This is relevant when considering that a sizable percentage of the voters only decides in the last few days~\cite{Hopmann2010}.
%this suggests that the media increasingly treat these parties as legitimate and newsworthy actors within the broader European political field. 
Importantly, this trend was not specific to particular outlets or countries. When analyzing references per news source, we found that over 59\% of the websites mention the Radical Right more than any other political family, including parties more central to national political competition (e.g., currently in office), and regardless of the newspapers audience size, or publication frequency. We also observed a high number of mentions to Radical Right entities in outlets traditionally aligned with the center left, such as \textit{The Guardian} or \textit{Der Spiegel}, (in 45\% and 35\% of all mentions, respectively). This points to a structural bias rather than local or contextual factors.

Such high visibility can be analyzed alongside the other salience dimensions. Our study did not explicitly investigate the context of the mentions and it is possible that the full news piece could be only tangentially related to electoral politics or involve ambiguous affiliations. These are unlikely possibilities. Regarding the latter, we applied strict term-based filters and manually verified all included entities, restricting the dataset to actors affiliated with recognized European Parliament groups or officially running in 2024, minimizing false positives. As for prominence, we only counted mentions in the news titles and their respective URLs (which often mimic the title), and it can be easily argued that they serve as a strong proxy for importance. Thus, we are confident that the analyzed pieces are indeed about EU politics and, even if they mention other parties in the body of the text, they offer disproportionate notability to the Radical Right.
Our analysis also did not evaluate valence, that is, the tone of the news, nor other aspects such as language use, framing, and narrative context. As the different outlets are known to cover different political orientations, including left and center-left newspapers or public broadcasters, it is plausible that much of the observed coverage was neutral or even negative. Given the large number of pieces, in different languages, human coding would be very taxing, and full automatic classification is still unreliable~\cite{zhang_etal_2024_sentiment}. However, as mentioned, previous studies have demonstrated that even negative coverage can increase recognition, legitimacy, and ultimately, the electoral performance of Radical Right actors~\cite{deVreese2006_2, Vliegenthart2012, Murphy2018, Dennison2019, Mancosu2021}. %For example, research conducted in Belgium, Germany, and the Netherlands~\cite{Vliegenthart2012} has shown that anti-immigration parties gained both visibility and electoral success when the media increased coverage of issues such as migration and cultural identity -- issues these parties are commonly perceived to “own.” Similarly, a panel study in Germany~\cite{Gei2017} found that party visibility in the news was a stronger predictor of vote switching than coverage of issue ownership itself. 

The present study was not designed to determine the causal nature of the relationship between media salience and electoral outcomes, and our results should not be interpreted as evidence that media visibility mechanically translates into electoral success for specific parties. Even so, in contexts where domestically relevant actors are weak or absent (as is the case of Ireland), media attention to foreign party family members can still matter politically by shaping their perceived public support, normative expectations, and the broader ideological climate. Previous research suggests that such effects can arise even in the absence of national organizational counterparts~\cite{selvanathan2025farright,bohmelt2024antiimmigration}, and may in fact precede both organizational entry and subsequent electoral success. Also, it is clear that Radical Right entities have been gaining electoral ground: not that many years ago, there were almost no mentions to such forces in countries like Austria, Germany or Portugal; however, this attention is now very high and, in the period since the 2024 EU elections, right-wing parties secured first place in Austrian legislative elections, second in the Portuguese, while in Germany, the Alternative for Germany (AfD) emerged as the strongest party in its eastern states.

We note that this convergence in coverage, across countries and political contexts, possibly reflects deeper transformations in the digital news ecosystem. Traditional outlets now operate under the same attention-driven logic as social media platforms, prioritizing engagement metrics such as clicks and shares. In this context, the Radical Right’s communicative style -- direct, emotional, and conflict-oriented -- fits the dynamics of visibility particularly well. 

%One possible explanation is that, as mainstream news media moves online, they interact with their viewers and readership through mechanisms that are equally sensitive to the pressures of the attention economy: views, "likes", comments, etc. In other words, editorial choices that favor engagement are likely to amplify the discourse that drives more engagement. 

Moreover, across Europe, this shift coincides with reports of circulation declines, significant reductions in the number of media outlets, and stagnation of digital signatures. Together, these forces create strong incentives to favor content that attracts attention. Editorial decisions that respond to audience analytics are likely to amplify the discourse that drives more engagement.
An alternative explanation, not mutually exclusive, is that efforts to demonstrate pluralism or counter accusations of bias have led some outlets to “overcompensate,” unintentionally amplifying these actors’ visibility.

In either case, claims of systematic censorship by Radical Right parties (and of a preference of the media towards the Radical Left), are difficult to sustain. On the contrary, these findings show that such actors are among the most visible in the European media landscape. This raises broader questions about pluralism, editorial responsibility, and how the distribution of media attention should be monitored and regulated in the digital age.

Future research should combine cross-national visibility data with longitudinal measures of tone, issue framing, and audience engagement, to explore whether media coverage merely mirrors or actively accelerates Europe’s rightward shift.

\section{Acknowledgments}
We thank members of the Social Physics and Complexity (SPAC) group at LIP for comments and critical reading of the manuscript. This research was partially funded by ERC Stg FARE (853566) and ERC PoC FARE-Audit (101100653), both to JGS, and by FCT PhD fellowship (2022.12547.BD) to ID.

\section{Acroynms}

\begin{acronym}
\acro{EP}{European Parliament}
\acro{EU}{European Union}
\acro{LLM}{Large Language Model}
\acro{MEPs}{Members of European Parliament}
%\acro{SEs}{Search Engines}
\end{acronym}

\printbibliography

\appendix
\setcounter{figure}{0}
\setcounter{table}{0}

\renewcommand{\thefigure}{A\arabic{figure}}
\renewcommand{\thetable}{A\arabic{table}}

\onecolumn

\section{Appendix}

\subsection{Keywords used to collect News on Media Cloud}

\begin{table}[ht!]

\caption{\textbf{Terms used to form queries used to collect news webpages from Media Cloud}.\\ Queries included all items in category 1 (bottom table), as well as all pairwise combinations of terms from categories 2 (middle table) and 3 (top table), joined with AND operator to account for non-contiguous mentions. EN\_IR corresponds to the English language, but considering the context in Ireland.}
\label{sup_table: mediacloud_queries} 

\centering
\small

\begin{tabular}{
>{\RaggedRight\arraybackslash}p{1.2cm}
>{\RaggedRight\arraybackslash}p{1.4cm}
>{\RaggedRight\arraybackslash}p{14.5cm}
}
\toprule
\textbf{Group} & \textbf{Language} & \textbf{Terms considered} \\
\midrule

EU - politics (G3) & DE & Debatte, Debatten, Ernannten, Kampagne, Kandidat, Kandidaten, Kandidatin, Kandidatinnen, Konvention, Kundgebung, Kundgebungen, Meinungsumfrage, Meinungsumfragen, Nominierten, Nominierter, Partei, Parteien, Parteitag, Spitzenkandidat, Spitzenkandidaten, Spitzenkandidatin, Stime, Stimme, Stimmen, Umfrage, Umfragen, Wahl, Wahlberechtigte, Wahlen, Wahlenberechtigte, Wahlenen, Wahlenkampf, Wahlenkampfveranstaltung, Wahlenkampfveranstaltungen, Wahlenstimme, Wahlenstimmen, Wähler, Wahlkampf, Wahlkampfveranstaltung, Wahlkampfveranstaltungen, Wahlstimme, Wahlstimmen \\[2pt] \hline

EU - politics (G3) & EN\_IR & Campaign, Candidate, Candidates, Convention, Debate, Debates, Election, Elections, Lead candidate, Lead candidates, Nominee, Nominees, Parties, Party, Poll, Polls, Rallies, Rally, Vote, Voter, Voters, Votes \\[2pt] \hline
EU - politics (G3) & PL & Debacie, Debat, Debat Wyborczych, Debata, Debata Wyborcza, Debatach, Debaty, Debaty Wyborczej, Głos, Głosować, Głosowania, Głosowanie, Głosy, Głosy wyborcze, Główni kandydaci, Główny kandydat, Kampania, Kampania Wyborcza, Kampanii, Kampanii Wyborczej, Kandydaci, Kandydat, Kandydata, Kandydatek, Kandydatem, Kandydatka, Kandydatką, Kandydatki, Kandydatów, Konwencja, Konwencja Wyborcza, Konwencji, Konwencji Wyborczej, Lider listy, Lidera listy, Liderek List, Liderka listy, Liderki List, Liderki listy, Liderów List, Liderów Listy, Liderzy List, Liderzy listy, Mianowana, Mianowane, Mianowani, Mianowany, Mianowanych, Mitingach Wyborczych, Mityng wyborczy, Mityngi wyborcze, Nominowana, Nominowane, Nominowani, Nominowany, Nominowanych, Partia, Partie, Partii, Sondaż, Sondaż Przedwyborczy, Sondaże, Sondaże Przedwyborcze, Sondażu, Sondaży Przedwyborczych, Spotkania Wyborcze, Spotkaniach Wyborczych, Spotkanie Wyborcze, Spotkaniu Wyborczym, W Sondażu Przedwyborczym, Wiec, Wiec Wyborczy, Wiecach Wyborczych, Wiece, Wiece Wyborcze, Wiecu Wyborczym, Wybór, Wybór Kandydata, Wybór Kandydatek, Wybór Kandydatki, Wybór Kandydatów, Wyborach, Wyborami, Wyborca, Wyborcy, Wyborczej, Wyborom, Wyborów, Wybory, Wybory Kandydatek, Wybory Kandydatów, Wyznaczeni, Wyznaczona, Wyznaczone, Wyznaczonego, Wyznaczonej, Wyznaczony, Wyznaczonych \\[2pt] \hline
EU - politics (G3) & PT & Cabeça de lista, Cabeças de lista, Campanha, Candidata, Candidatas, Candidato, Candidatos, Comício, Comícios, Convenção, Debate, Debates, Eleição, Eleições, Eleitor, Eleitores, Nomeada, Nomeadas, Nomeado, Nomeados, Partido, Partidos, Sondagem, Sondagens, Voto, Votos \\[2pt] \hline
    
\bottomrule
\end{tabular}

\end{table}

\begin{table}[ht!]
\centering
\small

\begin{tabular}{
>{\RaggedRight\arraybackslash}p{1.4cm}
>{\RaggedRight\arraybackslash}p{1.4cm}
>{\RaggedRight\arraybackslash}p{14cm}
}
\toprule
\textbf{Group} & \textbf{Language} & \textbf{Terms considered} \\
\midrule

General elections (G2) &  DE & Abstimmung in der EU, EU-Abgeordnete, EU-abgeordneten, EU-Abgeordneter, EU-Abstimmung, EU-kandidat, EU-Parlament, EU-Parlamentes, EU-Partei, EU-Parteien, Europaabgeordnete, Europa-abgeordnete, Europaabgeordneten, Europa-abgeordneten, Europaabgeordneter, Europa-abgeordneter, Europäische, Europäische Abstimmungen, Europäische Partei, Europäische Parteien, Europäischen Parlament, Europäischen Parteien, Europäisches Parlament, Europakandidat, Europa-kandidat, Europaparlament, Europa-Parlament, Europaparlamentes, Europa-Parlamentes, EU-Stimme, EU-Stimmen, Mitglied des Europäischen Parlaments, Mitglieder des Europäischen Parlaments \\[2pt] \hline
General elections (G2) &  EN\_IR & E.U. Votes, EU parliament, EU parties, EU party, EU vote, European, European deputies, European deputy, European parliament, European parties, European party, European votes, European voting, Member of the European Parliament, Members of the European Parliament, MEP, MEPs, Vote in the EU, Voting in the EU \\[2pt] \hline
%General elections (G2) &  EN\_US & Biden, Congress, Democrat, Democratic Party, Democrats, Electoral College, Electors, Harris, House of representatives, Member of the house, Presidential, Presidential debate, Presidentials, Republican, Republican Party, Republicans, Senate, Senator, Trump, Vance, Waltz, White house \\ \hline
General elections (G2) &  PL & Członek Parlamentu Europejskiego, Członkiń Parlamentu Europejskiego, Członkini Parlamentu Europejskiego, Członkinie Parlamentu Europejskiego, Członków Parlamentu Europejskiego, Członkowie Parlamentu Europejskiego, Eurodeputowana, Eurodeputowane, Eurodeputowanej, Eurodeputowani, Eurodeputowany, Eurodeputowanych, Eurodeputowanymi, Europarlamencie, Europarlament, Europarlamentu, Europejska, Europejska partia, Europejskie, Europejskie partie, Europoseł, Europosła, Europosłanka, Europosłanki, Europosłów, Europosłowie, Głos w UE, Głosowania europejskie, Głosowanie europejskie, Głosowanie w UE, Głosy w EU, Parlamencie Europejskim, Parlament Europejski, Parlament UE, Parlamentu Europejskiego, Partia Europejska, Partia UE, Partie Europejskie, Partie UE, Partii Europejskich, Partii Europejskiej, Partii UE, Poseł do Parlamentu Europejskiego, Posłanek do Parlamentu Europejskiego, Posłanka do Parlamentu Europejskiego, Posłanki do Parlamentu Europejskiego, Posłów do Parlamentu Europejskiego \\[2pt] \hline
General elections (G2) &  PT & Deputada europeia, Deputadas europeias, Deputado europeu, Deputados europeus, Eurodeputada, Eurodeputadas, Eurodeputado, Eurodeputados, Europeias, Membro do Parlamento europeu, Membros do Parlamento europeu, Parlamento europeu, Partido europeu, Partidos europeus, Votação europeia, Votação na UE, Votações europeias, Voto na UE \\[2pt] \hline
    
\bottomrule
\end{tabular}
\end{table}

\begin{table}[ht!]
\centering
\small

\begin{tabular}{
>{\RaggedRight\arraybackslash}p{1cm}
>{\RaggedRight\arraybackslash}p{1.4cm}
>{\RaggedRight\arraybackslash}p{14.5cm}
}
\toprule
\textbf{Group} & \textbf{Language} & \textbf{Terms considered} \\
\midrule

EU-elections (G1) & DE & EU-Parlament Wahlen, EU-Parlaments Wahlen, Europäische Parlamentswahlen, Europawahl, Europa-Wahl, Europawahlen, Europa-Wahlen, Europawahlkampf, Europa-Wahlkampf, Europawahlkampfauftritt, Europa-Wahlkampfauftritt, EU-Wahl, EU-Wahlen, EU-Wahlkampf, EU-Wahlkampfauftritt, Wahl für das EU-Parlament, Wahl für das Europäische Parlament, Wahl für das PE, Wahl für die EU, Wahlen der Europäischen Union, Wahlen des EU-Parlaments, Wahlen des Europäischen Parlaments, Wahlen EU, Wahlen EU-Parlament, Wahlen EU-Parlaments, Wahlen Europäische Parlament, Wahlen Europäische Parlaments, Wahlen Europäischen Parlament, Wahlen Europäischen Parlaments, Wahlen Europäischen Union, Wahlen für das EU, Wahlen für das EU-Parlament, Wahlen für das EU-Parlaments, Wahlen für das Europäische Parlament, Wahlen für das Europäische Parlaments, Wahlen für das Europäischen Parlament, Wahlen für das Europäischen Parlaments, Wahlen für das Europäischen Union, Wahlen für das PE, Wahlen für des EU, Wahlen für des EU-Parlament, Wahlen für des EU-Parlaments, Wahlen für des Europäische Parlament, Wahlen für des Europäische Parlaments, Wahlen für des Europäischen Parlament, Wahlen für des Europäischen Parlaments, Wahlen für des Europäischen Union, Wahlen für des PE, Wahlen für die EU, Wahlen für die EU-Parlament, Wahlen für die EU-Parlaments, Wahlen für die Europäische Parlament, Wahlen für die Europäische Parlaments, Wahlen für die Europäische Union, Wahlen für die Europäischen Parlament, Wahlen für die Europäischen Parlaments, Wahlen für die Europäischen Union, Wahlen für die PE, Wahlen für EU, Wahlen für EU-Parlament, Wahlen für EU-Parlaments, Wahlen für Europäische Parlament, Wahlen für Europäische Parlaments, Wahlen für Europäischen Parlament, Wahlen für Europäischen Parlaments, Wahlen für Europäischen Union, Wahlen für PE, Wahlen in der EU, Wahlen in der EU-Parlament, Wahlen in der EU-Parlaments, Wahlen in der Europäische Parlament, Wahlen in der Europäische Parlaments, Wahlen in der Europäischen Parlament, Wahlen in der Europäischen Parlaments, Wahlen in der Europäischen Union, Wahlen in der PE, Wahlen in EU, Wahlen in EU-Parlament, Wahlen in EU-Parlaments, Wahlen in Europäische Parlament, Wahlen in Europäische Parlaments, Wahlen in Europäischen Parlament, Wahlen in Europäischen Parlaments, Wahlen in Europäischen Union, Wahlen in PE, Wahlen PE, Wahlen zum EU, Wahlen zum EU-Parlament, Wahlen zum EU-Parlaments, Wahlen zum Europäische Parlament, Wahlen zum Europäische Parlaments, Wahlen zum Europäischen Parlament, Wahlen zum Europäischen Parlaments, Wahlen zum Europäischen Union, Wahlen zum PE, Wahlen zur EU, Wahlen zur EU-Parlament, Wahlen zur EU-Parlaments, Wahlen zur Europäische Parlament, Wahlen zur Europäische Parlaments, Wahlen zur Europäischen Parlament, Wahlen zur Europäischen Parlaments, Wahlen zur Europäischen Union, Wahlen zur PE, Wahlenen EU-Parlament, Wahlenen EU-Parlaments, Wahlenen Europäische Parlament, Wahlenen Europäische Parlaments, Wahlenen Europäischen Union \\[2pt] \hline
EU-elections (G1) & EN\_IR & Election for the EU, Election for the European Parliament, Elections for the EU, Elections for the European Parliament, Elections to the EU Parliament, Elections to the European Parliament, EP election, EP elections, EU election, EU elections, EU parliament elections, European election, European elections, European Parliament elections \\[2pt] \hline
%EU-elections (G1) & EN\_US & 2024 election, 2024 elections, Congressional election, Congressional elections, House election, House elections, House of representatives election, House of representatives elections, Presidential election, Presidential elections, Senate election, Senate elections, U.S. election, U.S. elections, U.S. house election, U.S. house elections, United State election, United State elections, United States election, United States elections, US election, US elections, US house election, US house elections \\ \hline
 EU-elections (G1) & PL & Eurowyborach, Eurowyborami, Eurowyborów, Eurowybory, Wyborach do europarlamentu, Wyborach do Parlamentu Europejskiego, Wyborach do parlamentu UE, Wyborach do PE, Wyborach do UE, Wyborach europejskich, Wyborami do europarlamentu, Wyborami do Parlamentu Europejskiego, Wyborami do parlamentu UE, Wyborami do PE, Wyborami do UE, Wyborami europejskimi, Wyborczej do europarlamentu, Wyborczej do Parlamentu Europejskiego, Wyborczej do parlamentu UE, Wyborczej do PE, Wyborczej do UE, Wyborów do europarlamentu, Wyborów do Parlamentu Europejskiego, Wyborów do parlamentu UE, Wyborów do PE, Wyborów do UE, Wyborów europejskich, Wybory do europarlamentu, Wybory do europy, Wybory do Parlamentu Europejskiego, Wybory do Parlamentu UE, Wybory do PE, Wybory do UE, Wybory do Unii Europejskiej, Wybory europejskie, Wybory eurpejskie \\[2pt] \hline
EU-elections (G1) & PT & Eleição da UE, Eleição europeia, Eleição para a UE, Eleição para o Parlamento europeu, Eleições ao PE, Eleições da UE, Eleições europeias, Eleições para a UE, Eleições para o Parlamento europeu \\[2pt] \hline
    
\bottomrule
\end{tabular}

\end{table}

\begin{table}[ht!]

\caption{\textbf{Terms used to identify Irish local elections and the second round of Polish local elections, applied in the exclusion mechanism.} EN\_IR corresponds to the English language, but considering the context in Ireland.}
\label{sup_table:mediacloud_queries_exclusion}

\begin{tabular}{
>{\RaggedRight\arraybackslash}p{1.7cm}
>{\RaggedRight\arraybackslash}p{15cm}
}
\toprule
\textbf{Language} & \textbf{Terms considered} \\
\midrule \hline

EN\_IR & General election, General elections, Local election, Local elections \\ \hline
PL & Wyborami samorządowymi, Wybory samorządowe \\ \hline
\end{tabular}%

\end{table}

\FloatBarrier
\subsection{Prompts used on ChatGPT for extraction and mapping tasks}

\begin{table}[ht!]

\caption{\textbf{Exact prompt for ChatGPT-4o used during political entities extraction, per country}}
\label{sup_table:prompt_political_entities_extraction}

\begin{tabular}{
>{\RaggedRight\arraybackslash}p{1.7cm} 
>{\RaggedRight\arraybackslash}p{15cm} }
\toprule
\textbf{Country} & \textbf{Prompt} \\
\midrule
\hline
    Austria & Identify and return the political parties, politicians, and other mentions to the political leaning of the mentioned agents or content (e.g., ``extreme-right'' or ``socialists'') in each title and URL. The titles and URLs provided are more likely to come from news written in German or in English, since correspond to titles and URLs of the newspapers with more visualizations in Austria. The entities to be found should be related to the Austrian political environment. The title and URLs are from news from April 9th until June 9th 2024, so consider political information, such as the name of the prime minister, from that period. The result should be given as a JSON object with 3 keys: ``title'' of type string with the value in the title column in the provided table, ``url'' of type string with the value in the url column in the provided table, ``parties'' of type dict with keys corresponding to the direct transcription of the found parties and values corresponding to the respective official name if acronym returned for instance, or null if not possible to infer anything, e.g. {``ÖVP'': ``Austrian People's Party - Austria''}) and ``politicians\_othermentions'' of type dict with key corresponding to the direct transcription of the found politicians and other mentions and the value corresponding with their associated party, e.g.: {``Chancellor of Austria'':``Austrian People's Party - Austria''}.     \\ \hline
     Germany & Identify and return the political parties, politicians, and other mentions to the political leaning of the mentioned agents or content (e.g., ``extreme-right'' or ``socialists'') in each title and URL. The titles and URLs provided are more likely to come from news written in German, since correspond to titles and URLs of the newspapers with more visualizations in Germany. The entities to be found should be related to the German political environment. The result should be given as a JSON object with 3 keys: ``title'' of type string with the value in the title column in the provided table, ``url'' of type string with the value in the url column in the provided table,``parties'' of type dict with keys corresponding to the direct transcription of the found parties and values corresponding to the respective official name if acronym returned for instance, or null if not possible to infer anything, e.g. {``SPD'': ``Social Democratic Party of Germany - Germany''})
 and ``politicians\_othermentions'' of type dict with key corresponding to the direct transcription of the found politicians and other mentions and the value corresponding with their associated party, e.g.: {``Chancellor of Germany'':``Social Democratic Party of Germany - Germany''}. \\ \hline
    Ireland & Identify and return the political parties, politicians, and other mentions to the political leaning of the mentioned agents or content (e.g., ``extreme-right'' or ``socialists'') in each title and URL. The titles and URLs provided are more likely to come from news written in Irish, since correspond to titles and URLs of the newspapers with more visualizations in Ireland. The entities to be found should be related to the Irish political environment. The title and URLs are from news from April 9th until June 9th 2024, so consider political information, such as the name of the prime minister, from that period.The result should be given as a JSON object with 3 keys: ``title`` of type string with the value in the title column in the provided table, ``url'' of type string with the value in the url column in the provided table, ``parties'' of type dict with keys corresponding to the direct transcription of the found parties and values corresponding to the respective official name if acronym returned for instance, or null if not possible to infer anything, e.g. {``Shinners'': ``Sinn Feín - Ireland''}) and ``politicians\_othermentions'' of type dict with key corresponding to the direct transcription of the found politicians and other mentions and the value corresponding with their associated party, e.g.: {``Taoiseach'':``Fine Gael - Ireland''}.\\ \hline
  
\end{tabular}%

\end{table}

\begin{table}[]
    \centering
\begin{tabular}{
>{\RaggedRight\arraybackslash}p{1.7cm} 
>{\RaggedRight\arraybackslash}p{15cm} }
\toprule
\textbf{Country} & \textbf{Prompt} \\
\midrule
\hline
      Poland & Identify and return the political parties, politicians, and other mentions to the political leaning of the mentioned agents or content (e.g., ``extreme-right'' or ``socialists'') in each title and URL. The titles and URLs provided are more likely to come from news written in Polish, since correspond to titles and URLs of the newspapers with more visualizations in Poland. The entities to be found should be related to the Polish political environment. The result should be given as a JSON object with 3 keys: ``title'' of type string with the value in the title column in the provided table, ``url'' of type string with the value in the url column in the provided table, ``parties'' of type dict with keys corresponding to the direct transcription of the found parties and values corresponding to the respective official name if acronym returned for instance, or null if not possible to infer anything, e.g. {``Civic Platform'': ``Civic Platform of the Republic of Poland - Poland''})
 and ``politicians\_othermentions'' of type dict with key corresponding to the direct transcription of the found politicians and other mentions and the value corresponding with their associated party, e.g.: {``premier polski'':``Civic Platform of the Republic of Poland - Poland''}. \\ \hline
    Portugal & Identify and return the political parties, politicians, and other mentions to the political leaning of the mentioned agents or content (e.g., ``extreme-right'' or ``socialists'') in each title and URL. The titles and URLs provided are more likely to come from news written in Portuguese, since correspond to titles and URLs of the newspapers with more visualizations in Portugal. The entities to be found should be related to the Portuguese political environment. The title and URLs are from news from April 9th until June 9th 2024, so consider political information, such as the name of the prime minister, from that period. The result should be given as a JSON object with 3 keys: ``title'' of type string with the value in the title column in the provided table, ``url'' of type string with the value in the url column in the provided table, ``parties'' of type dict with keys corresponding to the direct transcription of the found parties and values corresponding to the respective official name if acronym returned for instance, or null if not possible to infer anything, e.g. {``socialists'': ``Partido Socialista - Portugal''}
 and ``politicians\_othermentions'' of type dict with key corresponding to the direct transcription of the found politicians and other mentions and the value corresponding with their associated party, e.g.: {``primeiro ministro de Portugal'':``Partido Social Democrata - Portugal''}.  \\ \hline
\end{tabular}
\end{table}

\begin{figure*}[ht!]
    \centering
    \caption{\textbf{Prompt given to ChatGPT through the API, to identify the political party(ies) or politican associated with a certain political spectrum mention in the URL and Headline (``the left in Germany'').}}
    \label{sup_fig:identify_party}
    \includegraphics[width=1\textwidth]{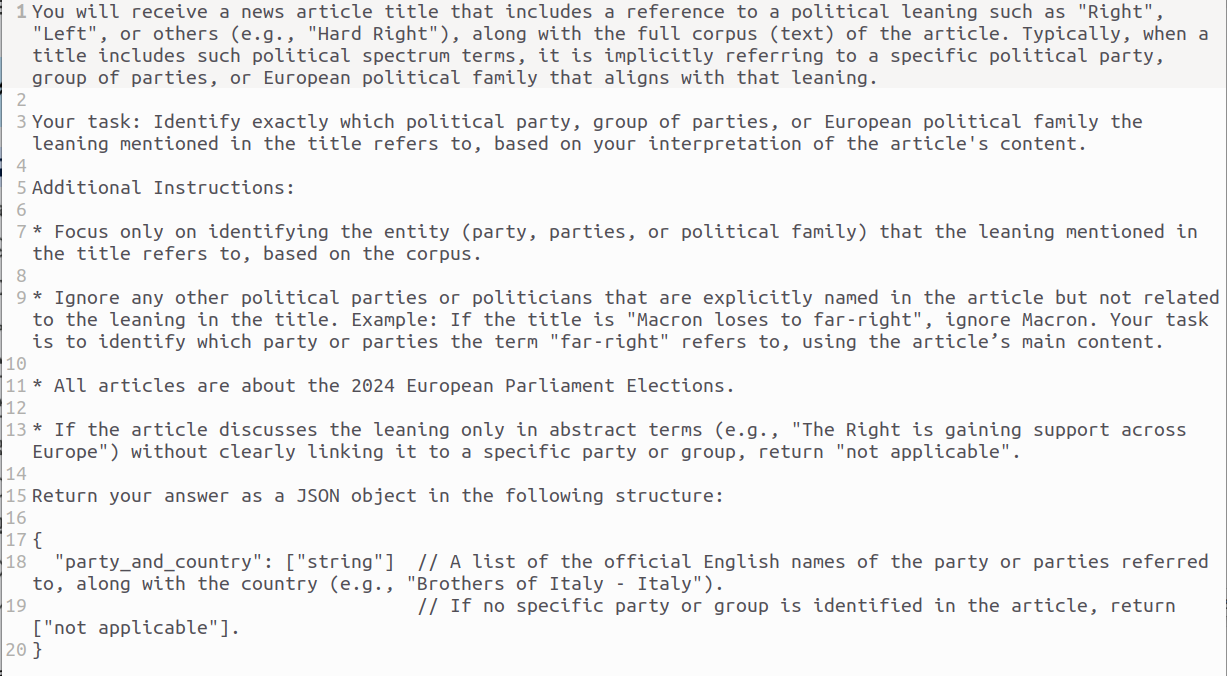}
    
\end{figure*}

\clearpage
\subsection{National Parties and respective classification}\label{national_parties_and_leaning}

\small
\setlength{\tabcolsep}{4pt} % tighter column spacing

\begin{longtable}{>{\RaggedRight\arraybackslash}p{9cm} 
                  >{\RaggedRight\arraybackslash}p{3cm}
                  >{\RaggedRight\arraybackslash}p{3cm}
                  >{\RaggedRight\arraybackslash}p{2cm}}
\caption{\textbf{Valid political parties in all headlines and URLs analyzed, their respective Chapel Hill political classification, the European Political group that they integrate with groups formed in 2024 mapped to the single Identity and Democracy (ID) in 2019.} The last column is our final leaning classification, mapped according to the Chapel Hill political classification.}
\label{tab:national_parties_leaning} \\
\toprule
\textbf{Party Name – Country} & 
\textbf{Chapel Hill political classification} &
\textbf{EU group} &
\textbf{Final group classification} \\
\midrule
\endfirsthead
\toprule
\textbf{Party Name – Country} & 
\textbf{Chapel Hill political classification} &
\textbf{EU group} &
\textbf{Final group classification} \\
\midrule \midrule
\endhead
\rowcolors{2}{gray!20}{white} 
% ---- Paste your data below this line ----
Action of Dissatisfied Citizens 2011 – Czech Republic & - & Patriots (ID) & Rad Right \\ \hline
Alliance 90 / Greens – Germany & green & G/EFA & Greens \\ \hline
Alliance for the Unity of Romanians – Romania & radrt & ECR & Rad Right \\ \hline
Alternative for Germany – Germany & radrt & Soverign (ID) & Rad Right \\ \hline
Austrian People's Party – Austria & cd & EPP & Main Right \\ \hline
Bloc of the Left – Portugal & radleft & The Left & Rad Left \\ \hline
Brothers of Italy – Italy & radrt & ECR & Rad Right \\ \hline
Centre for Poland – Poland &  & EPP and Renew & Main Right \\ \hline
Centre Party – Sweden & agrarian/centre & Renew & Main Right \\ \hline
Christian Democratic Appeal – Netherlands & cd & EPP & Main Right \\ \hline
Christian Democratic People's Party – Hungary & radrt & Patriots (ID) & Rad Right \\ \hline
Christian Democratic Union / Christian Social Union – Germany & cd & EPP & Main Right \\ \hline
Christian Democrats – Sweden & cd & EPP & Main Right \\ \hline
Christian Social People's Party – Luxembourg & - & EPP & Main Right \\ \hline
Christian Social Union – Germany & cd & EPP & Main Right \\ \hline
Christian Social Union in Bavaria – Germany & cd & - & Main Right \\ \hline
Citizens for European Development of Bulgaria – Bulgaria & con & EPP & Main Right \\ \hline
Civic Democratic Party – Czech Republic & con & ECR & Main Right \\ \hline
Civic Platform – Poland & cd & EPP & Main Right \\ \hline
Coalition for the Renewal of the Republic –– Liberty and Hope – Poland & radrt & Soverign (ID) & Rad Right \\ \hline
Coalition of the Radical Left – Greece & radleft & The Left & Rad Left \\ \hline
Conservative People's Party – Estonia & radrt & Patriots (ID) & Rad Right \\ \hline
Croatian Democratic Union – Croatia & cd & EPP & Main Right \\ \hline
Danish People's Party – Denmark & radrt & Patriots (ID) & Rad Right \\ \hline
Danish Social Liberal Party – Denmark & lib & Renew & Main Right \\ \hline
Democratic Alliance – Portugal & lib and con & EPP & Main Right \\ \hline
Democratic alliance – Portugal & lib and con & EPP & Main Right \\ \hline
Democratic and Social Centre –– People's Party – Portugal & con & EPP & Main Right \\ \hline
Democratic Convergence | Together for Catalonia – Spain & reg & Renew & Main Right \\ \hline
Democratic Left Alliance – Poland & soc & S\&D & Main Left \\ \hline
Democratic Party – Italy & soc & S\&D & Main Left \\ \hline
Democratic Party – Luxembourg & - & Renew & Main Right \\ \hline
Democrats 66 – Netherlands & lib & Renew & Main Right \\ \hline
Direction –– Social Democracy – Slovakia & soc & noattachedmembers & Main Left \\ \hline
Ecological Democratic Party – Germany & - & EPP & Main Right \\ \hline
Ecology Party –– Greens – Portugal & radleft & The Left & Rad Left \\ \hline
Enough – Portugal & radrt & Patriots (ID) & Rad Right \\ \hline
Estonian Reform Party – Estonia & lib & Renew & Main Right \\ \hline
Farmer-Citizen Movement – Netherlands & - & EPP & Main Right \\ \hline
Federation of the Greens – Italy & green & Greens & Greens \\ \hline
Fianna Fail – Ireland & con & Renew & Main Right \\ \hline
Fidesz –– Hungarian Civic Union – Hungary & radrt & Patriots (ID) & Rad Right \\ \hline
Fine Gael (Family of the Irish) – Ireland & cd & EPP & Main Right \\ \hline
Finnish Party | True Finns – Finland & radrt & ECR & Rad Right \\ \hline
Flemish Christian Peoples Party | Christian Democrats \& Flemish – Belgium & cd & EPP & Main Right \\ \hline
Flemish Interest – Belgium & radrt & Patriots (ID) & Rad Right \\ \hline
Forum for Democracy – Netherlands & radrt & noattachedmembers & Rad Right \\ \hline
Free Democratic Party – Germany & lib & Renew & Main Right \\ \hline
Free Voters – Germany & - & Renew & Main Right \\ \hline
Freedom Party of Austria – Austria & radrt & Patriots (ID) & Rad Right \\ \hline
Galician Nationalist Bloc – Spain & reg & Greens & Greens \\ \hline
Go Italy –– The People of Freedom – Italy & con & EPP & Main Right \\ \hline
Greek Solution – Greece & radrt & ECR & Rad Right \\ \hline
Greek Solution – Greece & radrt & ECR & Rad Right \\ \hline
Green Europe – Italy & green & Greens & Greens \\ \hline
Green League – Finland & green & Greens & Greens \\  \hline
Green Party – Ireland & green & Greens & Greens \\ \hline
Greens – Sweden & green & Greens & Greens \\ \hline
Homeland Union – Croatia & radrt & ECR & Rad Right \\ \hline
Homeland Union – Lithuania & con & EPP & Main Right \\ \hline
Independent Ireland – Ireland & - & Renew & Main Right \\ \hline
Independents 4 Change – Ireland & radleft & The Left & Rad Left \\ \hline
Italy Alive – Italy & lib & Renew & Main Right \\ \hline
Jobbik Movement for a Better Hungary – Hungary & radrt & noattachedmembers & Rad Right \\ \hline
Labour Party – Ireland & soc & S\&D & Main Left \\ \hline
Labour Party – Netherlands & soc & S\&D & Main Left \\ \hline
Labour Union – Poland & cd & S\&D & Main Left \\ \hline
Latvia First – Latvia &  & Patriots (ID) & Rad Right \\ \hline
Law and Justice – Poland & radrt & ECR & Rad Right \\ \hline
Left – Italy & soc & left and greens & Main Left \\ \hline
Left Together – Poland & radleft & The Left & Rad Left \\ \hline
Liberal Initiative – Portugal & lib & Renew & Main Right \\ \hline
Lithuanian Social Democratic Party – Lithuania & soc & S\&D & Main Left \\ \hline
Livre – Portugal & green & Greens & Greens \\ \hline
Luxembourg Socialist Workers' Party – Luxembourg & soc & S\&D & Main Left \\ \hline
Malta Labour Party – Malta & soc & S\&D & Main Left \\ \hline
Moderate Party – Sweden & con & EPP & Main Right \\ \hline
Modern – Poland & cd & Renew & Main Right \\ \hline
National Alliance – Latvia & radrt & ECR & Rad Right \\ \hline
National Coalition Party – Finland & con & EPP & Main Right \\ \hline
National Liberal Party – Romania & lib & EPP & Main Right \\ \hline
National Movement – Poland & radrt & Patriots (ID) & Rad Right \\ \hline
National Rally – France & radrt & Patriots (ID) & Rad Right \\ \hline
Nationalist Party – Malta & con & EPP & Main Right \\ \hline
NEOS –– The New Austria – Austria & lib & Renew & Main Right \\ \hline
New Democracy – Greece & con & EPP & Main Right \\ \hline
New Flemish Alliance – Belgium & reg & ECR & Rad Right \\ \hline
New Left – Poland & soc & S\&D & Main Left \\ \hline
New Social Contract – Netherlands & cd & EPP & Main Right \\ \hline
North League – Italy & radrt & Patriots (ID) & Rad Right \\ \hline
Our Homeland Movement – Hungary & radrt & Soverign (ID) & Rad Right \\ \hline
Party for Animals and Nature – Portugal & green & Greens & Greens \\ \hline
Party for Freedom – Netherlands & radrt & Patriots (ID) & Rad Right \\ \hline
Party of Liberty and Progress | Flemish Liberals and Democrats – Belgium & lib & Renew & Main Right \\ \hline
PASOK - Greece & soc & pasdeu & Main Left \\ \hline
PDS | The Left - Germany & radleft & left & Rad Left \\ \hline
People's Party - Spain & com & EPP & Main Right \\ \hline
People's Party for Freedom and Democracy - Netherlands & lib & Renew & Main Right \\ \hline
Peoples Association -- Golden Dawn - Greece &  & noattachedmembers & Rad Right \\ \hline
Podemos - Spain & radleft & left & Rad Left \\ \hline
Poland 2050 - Poland & lib & Renew & Main Right \\ \hline
Polish Initiative - Poland & cd & EPP & Main Right \\ \hline
Polish People's Party - Poland & agrarian/centre & EPP & Main Right \\ \hline
Political Reformed Party - Netherlands & confessional & ECR & Rad Right \\ \hline
Popular Monarchist Party - Portugal & lib and con & EPP & Main Right \\ \hline
Portuguese Communist Party - Portugal & radleft & left & Rad Left \\ \hline
Pro Patria Union - Estonia & con & EPP & Main Right \\ \hline
Progressive Slovakia - Slovakia & lib & Renew & Main Right \\ \hline
Reason and Justice - Germany & radleft & noattachedmembers & Rad Left \\ \hline
Reconquest - France & radrt & Soverign (ID) & Rad Right \\ \hline
Reformist Movement - Belgium & lib & Renew & Main Right \\ \hline
Republican Left of Catalonia - Spain & reg & Greens & Greens \\ \hline
Respect and Freedom Party - Hungary & cd & EPP & Main Right \\ \hline
Revival - Bulgaria & radrt & Soverign (ID) & Rad Right \\ \hline
Sinn Fein - Ireland & reg & left & Rad Left \\ \hline
Slovenian Democratic Party - Slovenia & con & EPP & Main Right \\ \hline
Social Democratic Party - Portugal & lib & EPP & Main Right \\ \hline
Social Democratic Party - Romania & soc & S\&D & Main Left \\ \hline
Social Democratic Party of Austria - Austria & soc & S\&D & Main Left \\ \hline
Social Democratic Party of Croatia - Croatia & soc & S\&D & Main Left \\ \hline
Social Democratic Party of Germany - Germany & soc & S\&D & Main Left \\ \hline
Social Democrats - Denmark & soc & S\&D & Main Left \\ \hline
Social Democrats - Sweden & soc & S\&D & Main Left \\ \hline
Socialist Party - France & soc & S\&D & Main Left \\ \hline
Socialist Party - Portugal & soc & S\&D & Main Left \\ \hline
Socialist Party [Francophone] - Belgium & soc & S\&D & Main Left \\ \hline
Socialist Peoples Party - Denmark & green & Greens & Greens \\ \hline
South Tyrol Peoples Party - Italy & reg & EPP & Main Right \\ \hline
Spanish Socialist Workers Party - Spain & soc & S\&D & Main Left \\ \hline
Spring - Poland & soc & S\&D & Main Left \\ \hline
Sweden Democrats - Sweden & radrt & ECR & Rad Right \\ \hline
The Greens - Netherlands & green & Greens & Greens \\ \hline
The Greens - Poland & cd & Greens & Greens \\ \hline
The Greens -- The Green Alternative - Austria & green & Greens & Greens \\ \hline
The Party is Over - Spain & radrt & noattachedmembers & Rad Right \\ \hline
The Republic Onwards! | Renaissance - France & lib & Renew & Main Right \\ \hline
Together Party - Poland & radleft & left & Rad Left \\ \hline
Unbowed France - France & radleft & left & Rad Left \\ \hline
Unified Democratic Coalition - Portugal & radleft & left & Rad Left \\ \hline
Union for a Popular Movement | The Republicans - France & con & EPP & Main Right \\ \hline
Union for French Democracy | Democratic Movement - France & lib & Renew & Main Right \\ \hline
United Poland - Poland & radrt & ECR & Rad Right \\ \hline
United We Can - Spain & radleft & left & Rad Left \\ \hline
Unity - Latvia & lib & EPP & Main Right \\ \hline
Victory - Greece & confessional & noattachedmembers & Rad Right \\ \hline
Voice - Social Democracy - Slovakia & soc & noattachedmembers & Main Left \\ \hline
Voice - Spain & radrt & Patriots (ID) & Rad Right \\ \hline
We Can - Spain & radleft & left & Rad Left \\ \hline
We Can! -- Political Platform - Croatia & green & Greens & Greens \\ \hline

\bottomrule

\end{longtable}

\clearpage
\subsection{Supplementary analysis of Political Mentions per Country}

\begin{figure*}[ht!]
    \centering
    \caption{\textbf{a) Frequency of the EU political groups per leaning (color) and country (row). b) Frequency of national parties (same country) and non-national ones (dashed section), per leaning (color) and country (row).}}
    \includegraphics[width=0.90\textwidth]{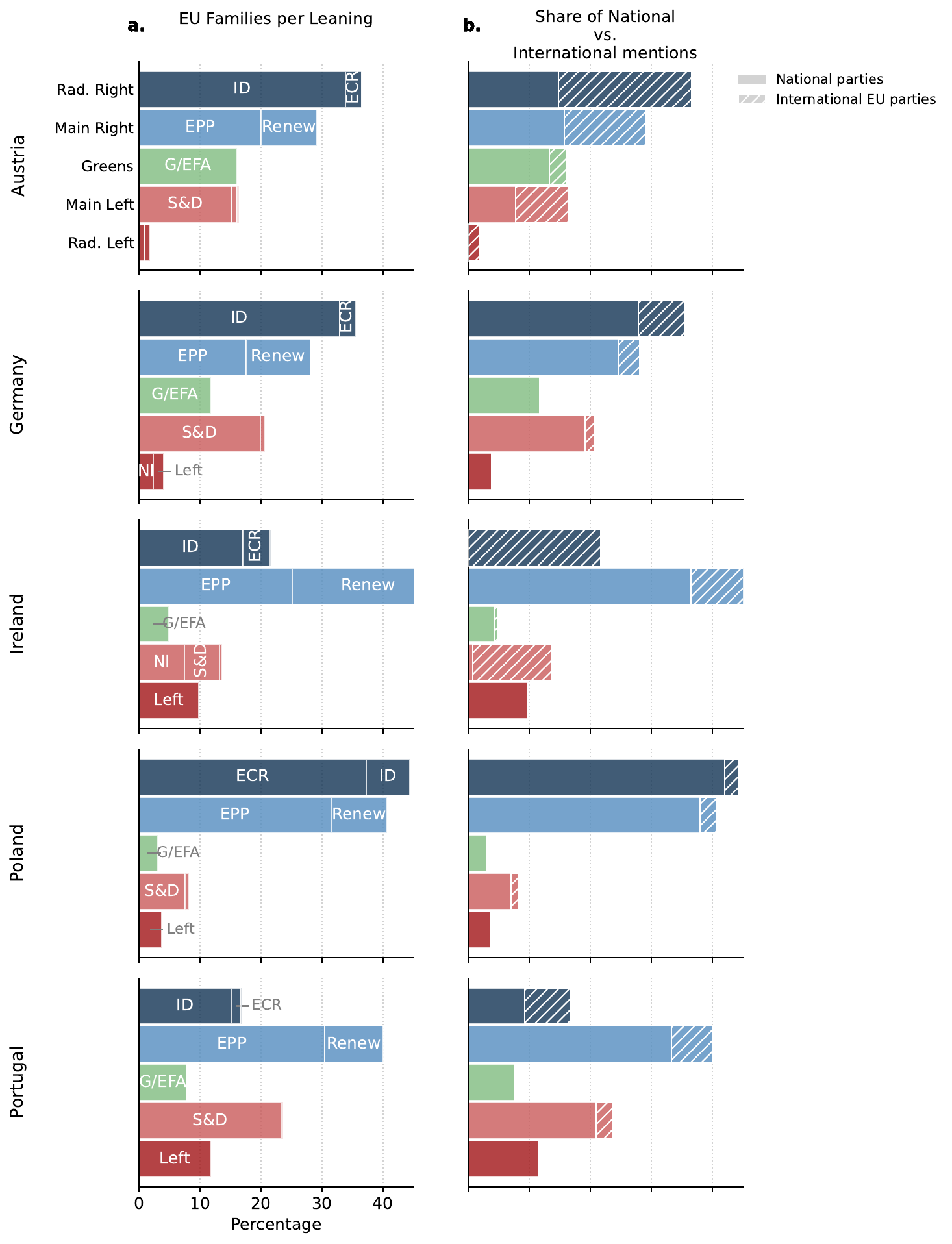}
    \label{fig:eu_parties_countries}
\end{figure*}

\begin{figure*}[ht!]
    \centering
    \caption{\textbf{Bars represent average frequency of exclusively national political entities, grouped by main ideological leaning and weighted by each country’s number of seats.} Triangles correspond to weighted outcomes of the previous election, crosses to weighted polling-based estimates, and dots to the current weighted allocation of seats among political groups.}
    \includegraphics[width=0.55\textwidth]{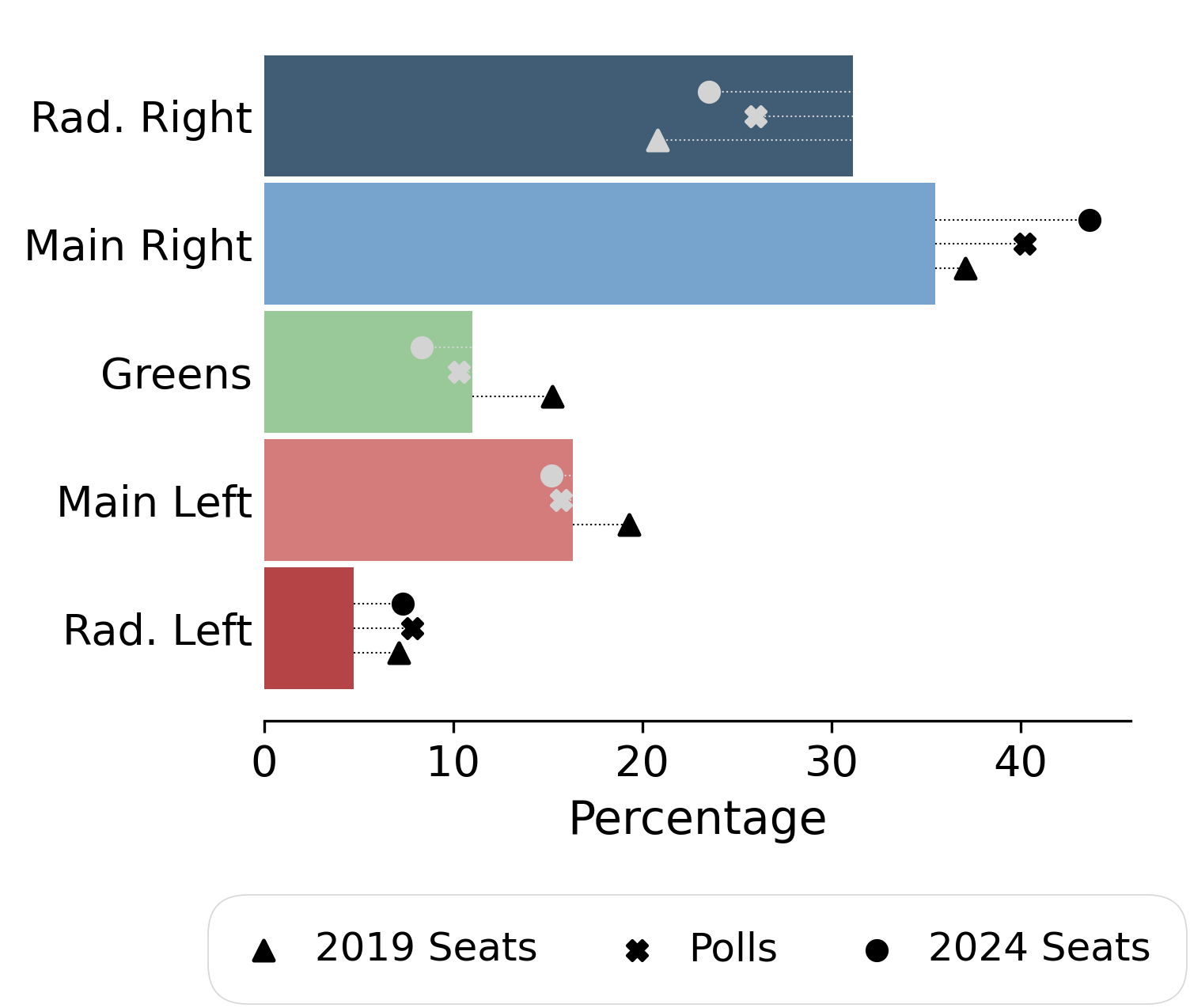}
    \label{fig:eu_parties_countries_weighted}
\end{figure*}

\begin{longtable}{|l|p{1.8cm}|p{4cm}|p{8cm}|p{0.65cm}|}
\caption{\textbf{For each country and main political category, the table lists the five most frequently mentioned political entities, along with their respective parties and share of mentions.} Mentions of the same political leaning but beyond the top five are grouped under the ``Other'' category, with their combined percentage reported.} \label{tab:top_entities} \\
\toprule
Country & Main Political Leaning & Entity Identified & Party - Country & \% \\
\midrule
\endfirsthead
\caption[]{Top political entities per country and EU political family, with percentages.} \\
\toprule
Country & Main Political Leaning & Entity Identified & Party - Country & \% \\
\midrule
\endhead
% \midrule
% \multicolumn{5}{r}{Continued on next page} \\
% \midrule
\endfoot
\bottomrule
\endlastfoot
\multirow{11}{*}{Austria} & \multirow{6}{*}{Rad. Left} & PDS | The Left - Germany & PDS | The Left - Germany & 33.3 \\ \cline{3-5}
& & Reason and Justice - Germany & Reason and Justice - Germany & 22.2 \\ \cline{3-5}
& & Sahra Wagenknecht & Reason and Justice - Germany & 17.8 \\ \cline{3-5}
& & Aminata Belli & PDS | The Left - Germany & 4.4 \\ \cline{3-5}
& & Anti-Ampel Coalition - Germany & PDS | The Left - Germany & 4.4 \\ \cline{3-5}
& & Other & PDS | The Left - Germany; Reason and Justice - Germany; Bloc of the Left - Portugal; left; We Can - Spain & 17.8 \\
\cline{2-5}
& \multirow{5}{*}{Main Left} & Social Democratic Party of Austria - Austria & Social Democratic Party of Austria - Austria & 30.0 \\ \cline{3-5}
& & Social Democratic Party of Germany - Germany & Social Democratic Party of Germany - Germany & 15.9 \\ \cline{3-5}
& & Olaf Scholz & Social Democratic Party of Germany - Germany & 7.9 \\ \cline{3-5}
& & Andreas Schieder & Social Democratic Party of Austria - Austria & 7.2 \\ \cline{3-5}
& & Robert Fico & Direction -- Social Democracy - Slovakia & 4.3 \\ \cline{3-5}
\multirow{13}{*}{Austria} & Main Left & Other & Social Democratic Party of Austria - Austria; Social Democratic Party of Germany - Germany; Social Democrats - Denmark; Left - Italy; Spanish Socialist Workers Party - Spain; Democratic Party - Italy; not applicable; pasdeu; Socialist Party - France; Democratic Left Alliance - Poland; Spring - Poland; Direction -- Social Democracy - Slovakia; Voice - Social Democracy - Slovakia; Hungarian Socialist Party - Hungary; Luxembourg Socialist Workers' Party - Luxembourg; Social Democratic Party - Romania; Socialist Party [Francophone] - Belgium & 34.6 \\
\cline{2-5}
& \multirow{6}{*}{Greens} & Greens - Austria & The Greens -- The Green Alternative - Austria & 34.2 \\ \cline{3-5}
& & Lena Schilling & The Greens -- The Green Alternative - Austria & 22.7 \\ \cline{3-5}
& & Alexander Van der Bellen & The Greens -- The Green Alternative - Austria & 10.8 \\ \cline{3-5}
& & Greens - Germany & Alliance 90 / Greens - Germany & 7.1 \\ \cline{3-5}
& & Ampel coalition - Germany & Alliance 90 / Greens - Germany & 3.9 \\ \cline{3-5}
& & Other & The Greens -- The Green Alternative - Austria; Alliance 90 / Greens - Germany; Left - Italy; not applicable & 21.2 \\ \cline{2-5}
& \multirow{6}{*}{Main Right} & Austrian People's Party - Austria & Austrian People's Party - Austria & 16.6 \\ \cline{3-5}
& & NEOS -- The New Austria - Austria & NEOS -- The New Austria - Austria & 9.3 \\ \cline{3-5}
& & Ursula von der Leyen & Christian Democratic Union - Germany & 8.3 \\ \cline{3-5}
& & Christian Democratic Union - Germany & Christian Democratic Union - Germany & 7.3 \\ \cline{3-5}
& & Emmanuel Macron & The Republic Onwards! | Renaissance - France & 6.4 \\ 
\cline{3-5}
&  & Other & Austrian People's Party - Austria; Christian Democratic Union - Germany; Free Democratic Party - Germany; NEOS -- The New Austria - Austria; Civic Platform - Poland; Croatian Democratic Union - Croatia; Go Italy -- The People of Freedom - Italy; Moderate Party - Sweden; National Coalition Party - Finland; National Liberal Party - Romania; New Democracy - Greece; Slovenian Democratic Party - Slovenia; Social Democratic Party - Portugal; Union for a Popular Movement | The Republicans - France; Christian Democratic Union / Christian Social Union - Germany; People's Party for Freedom and Democracy - Netherlands; Christian Social Union - Germany; Respect and Freedom Party - Hungary; Reformist Movement - Belgium; Christian Social People's Party - Luxembourg; Democratic Alliance - Portugal; South Tyrol Peoples Party - Italy; epp; The Republic Onwards! | Renaissance - France; Modern - Poland; Poland 2050 - Poland; Polish Initiative - Poland; Polish People's Party - Poland; Free Voters - Germany; Flemish Christian Peoples Party | Christian Democrats \& Flemish - Belgium; People's Party - Spain; Christian Democrats - Sweden & 52.1 \\ \cline{2-5}
Austria & \multirow{6}{*}{Rad. Right} & Alternative for Germany - Germany & Alternative for Germany - Germany & 32.3 \\ \cline{3-5}
& & Freedom Party of Austria - Austria & Freedom Party of Austria - Austria & 25.8 \\ \cline{3-5}
& & Harald Vilimsky & Freedom Party of Austria - Austria & 5.5 \\ \cline{3-5}
& & Herbert Kickl & Freedom Party of Austria - Austria & 4.3\\ \cline{3-5}
& & Giorgia Meloni & Brothers of Italy - Italy & 3.9 \\ \cline{3-5}
& & Other & National Rally - France; Alternative for Germany - Germany; Freedom Party of Austria - Austria; Fidesz -- Hungarian Civic Party / Christian Democratic People's Party - Hungary; Brothers of Italy - Italy; Party for Freedom - Netherlands; Voice - Spain; Fidesz -- Hungarian Civic Union - Hungary; Law and Justice - Poland; North League - Italy; Action of Dissatisfied Citizens 2011 - Czech Republic; Peoples Association -- Golden Dawn - Greece; Enough - Portugal; Forum for Democracy - Netherlands; Finnish Party | True Finns - Finland; Alliance for the Unity of Romanians - Romania; Jobbik Movement for a Better Hungary - Hungary; National Alliance - Latvia; Reconquest - France; Sweden Democrats - Sweden  & 28.2 \\ \cline{1-5}
\multirow{14}{*}{Germany} & \multirow{6}{*}{Rad.Left} & Sahra Wagenknecht & Reason and Justice - Germany & 30.9 \\ \cline{3-5}
& & PDS | The Left - Germany & PDS | The Left - Germany & 26.8 \\ \cline{3-5}
& & Reason and Justice - Germany & Reason and Justice - Germany & 22.8 \\ \cline{3-5}
& & Party of Sahra Wagenknecht & Reason and Justice - Germany & 5.7 \\ \cline{3-5}
&  & Bodo Ramelow & PDS | The Left - Germany & 3.3 \\ \cline{3-5}
& & Other & PDS | The Left - Germany; Coalition of the Radical Left - Greece; left; Bloc of the Left - Portugal; Left Together - Poland & 10.6 \\ \cline{2-5}
& \multirow{6}{*}{Main Left} & Social Democratic Party of Germany - Germany & Social Democratic Party of Germany - Germany & 41.7 \\ \cline{3-5}
& & Olaf Scholz & Social Democratic Party of Germany - Germany & 15.0 \\ \cline{3-5}
& & Ampel coalition - Germany & Social Democratic Party of Germany - Germany & 9.6 \\ \cline{3-5}
& & Nancy Faeser & Social Democratic Party of Germany - Germany & 5.0 \\ \cline{3-5}
& & Robert Fico & Direction -- Social Democracy - Slovakia & 3.6 \\ \cline{3-5}
& & Other & Social Democratic Party of Germany - Germany; Social Democrats - Denmark; Spanish Socialist Workers Party - Spain; S\&D (European Group); Democratic Party - Italy; Labour Party - Netherlands; Spring - Poland; Direction -- Social Democracy - Slovakia; Luxembourg Socialist Workers' Party - Luxembourg; New Left - Poland; Social Democratic Party of Croatia - Croatia & 25.0 \\ \cline{2-5}
& \multirow{2}{*}{Greens} & Greens - Germany & Alliance 90 / Greens - Germany & 51.9 \\ \cline{3-5}
& & Ampel coalition - Germany & Alliance 90 / Greens - Germany & 19.3 \\ \cline{3-5}
\multirow{14}{*}{Germany} & \multirow{4}{*}{Greens} & Alliance 90 / Greens - Germany & Alliance 90 / Greens - Germany & 7.2 \\ \cline{3-5}
& & Annalena Baerbock & Alliance 90 / Greens - Germany & 5.2 \\ \cline{3-5}
& & Robert Habeck & Alliance 90 / Greens - Germany & 4.7 \\ \cline{3-5}
& & Other & Alliance 90 / Greens - Germany; The Greens -- The Green Alternative - Austria; Federation of the Greens - Italy; Galician Nationalist Bloc - Spain; Republican Left of Catalonia - Spain & 11.6 \\ \cline{2-5}
& \multirow{6}{*}{Main Right} & Christian Democratic Union - Germany & Christian Democratic Union - Germany & 22.2 \\ \cline{3-5}
& & Free Democratic Party - Germany & Free Democratic Party - Germany & 11.1 \\ \cline{3-5}
& & Ursula von der Leyen & Christian Democratic Union - Germany & 9.2 \\ \cline{3-5}
& & Ampel coalition - Germany & Free Democratic Party - Germany & 8.0 \\ \cline{3-5}
& & Emmanuel Macron & The Republic Onwards! | Renaissance - France & 6.6 \\ \cline{3-5}

 & & Other & Free Democratic Party - Germany; Christian Democratic Union / Christian Social Union - Germany; Christian Social Union - Germany; Christian Democratic Union - Germany; Free Voters - Germany; Civic Platform - Poland; People's Party for Freedom and Democracy - Netherlands; Progressive Slovakia - Slovakia; Austrian People's Party - Austria; Respect and Freedom Party - Hungary; epp; Croatian Democratic Union - Croatia; Citizens for European Development of Bulgaria - Bulgaria; Democratic Convergence | Together for Catalonia - Spain; Reformist Movement - Belgium; Ecological Democratic Party - Germany; The Republic Onwards! | Renaissance - France; Modern - Poland; Christian Social People's Party - Luxembourg; Estonian Reform Party - Estonia; Danish Social Liberal Party - Denmark; Nationalist Party - Malta; Christian Democrats - Sweden; Poland 2050 - Poland; Polish People's Party - Poland; Christian Democratic Appeal - Netherlands & 42.9 \\ \cline{2-5}
& \multirow{6}{*}{Rad. Right} & Alternative for Germany - Germany & Alternative for Germany - Germany & 66.8 \\ \cline{3-5}
& & Maximilian Krah & Alternative for Germany - Germany & 4.4 \\ \cline{3-5}
& & Giorgia Meloni & Brothers of Italy - Italy & 3.2 \\ \cline{3-5}
& & Far-right - Germany & Alternative for Germany - Germany & 3.0 \\ \cline{3-5}
& & Marine Le Pen & National Rally - France & 2.9 \\ \cline{3-5}
& & Other & Alternative for Germany - Germany; Party for Freedom - Netherlands; Identity and Democracy; Fidesz -- Hungarian Civic Party / Christian Democratic People's Party - Hungary; Freedom Party of Austria - Austria; National Rally - France; Brothers of Italy - Italy; Law and Justice - Poland; Fidesz -- Hungarian Civic Union - Hungary; Voice - Spain; ecr; North League - Italy; Alliance for the Unity of Romanians - Romania; Flemish Interest - Belgium; Reconquest - France & 16.6 \\ \cline{1-5}
\multirow{22}{*}{Ireland} & \multirow{6}{*}{Rad. Left} & Sinn Fein - Ireland & Sinn Fein - Ireland & 72.6 \\ \cline{3-5}
& & Mary Lou McDonald & Sinn Fein - Ireland & 9.5 \\ \cline{3-5}
& & Opposition - Ireland & Independents 4 Change - Ireland & 4.8 \\ \cline{3-5}
& & Clare Daly & Independents 4 Change - Ireland & 3.6 \\ \cline{3-5}
& & Cynthia Ní Mhurchú & Sinn Fein - Ireland & 2.4 \\ \cline{3-5}
& & Other & Sinn Fein - Ireland & 7.1 \\ \cline{2-5}
& \multirow{6}{*}{Main Left} & Robert Fico & Direction -- Social Democracy - Slovakia & 53.0 \\ \cline{3-5}
& & Mette Frederiksen & Social Democrats - Denmark & 12.0 \\ \cline{3-5}
& & Olaf Scholz & Social Democratic Party of Germany - Germany & 6.0 \\ \cline{3-5}
& & Pedro Sanchez & Spanish Socialist Workers Party - Spain & 5.1 \\ \cline{3-5}
& & Ilaria Salis - Greens and Left Alliance - Italy & Left - Italy & 2.6 \\ \cline{3-5}
& & Other & Socialist Party - Portugal; Labour Party - Ireland; not applicable; Left - Italy; Social Democratic Party of Germany - Germany; Spanish Socialist Workers Party - Spain; pasdeu; Norwegian Labour Party - Norway; Malta Labour Party - Malta; Luxembourg Socialist Workers' Party - Luxembourg; Direction -- Social Democracy - Slovakia & 21.4 \\ \cline{2-5}
& \multirow{6}{*}{Greens} & Greens - Ireland & Green Party - Ireland & 46.2 \\ \cline{3-5}
& & Eamon Ryan & Green Party - Ireland & 15.4 \\ \cline{3-5}
& & Ilaria Salis - Greens and Left Alliance - Italy & Left - Italy & 7.7 \\ \cline{3-5}
& & Separatists - Spain & Galician Nationalist Bloc - Spain & 5.8 \\ \cline{3-5}
& & Separatists - Spain & Republican Left of Catalonia - Spain & 5.8 \\ \cline{3-5}
& & Other & not applicable; Green Europe - Italy; Alliance 90 / Greens - Germany; Green Party - Ireland & 19.2 \\ \cline{2-5}
& \multirow{6}{*}{Main Right} & Emmanuel Macron & The Republic Onwards! | Renaissance - France & 11.9\\ \cline{3-5}
& & Government - Ireland & Fianna Fail - Ireland & 9.2\\ \cline{3-5}
& & Government - Ireland & Fine Gael (Family of the Irish) - Ireland & 9.2 \\ \cline{3-5}
& & Leo Varadkar & Fine Gael (Family of the Irish) - Ireland & 8.6 \\ \cline{3-5}
& & Fine Gael (Family of the Irish) - Ireland & Fine Gael (Family of the Irish) - Ireland & 7.8 \\ \cline{3-5}
\multirow{7}{*}{Ireland}& & Other & Fianna Fail - Ireland; Fine Gael (Family of the Irish) - Ireland; Christian Democratic Union - Germany; People's Party for Freedom and Democracy - Netherlands; Independent Ireland - Ireland; Party of Liberty and Progress | Flemish Liberals and Democrats - Belgium; Democratic Convergence | Together for Catalonia - Spain; Austrian People's Party - Austria; Civic Platform - Poland; Croatian Democratic Union - Croatia; Go Italy -- The People of Freedom - Italy; Moderate Party - Sweden; National Coalition Party - Finland; National Liberal Party - Romania; New Democracy - Greece; Slovenian Democratic Party - Slovenia; Social Democratic Party - Portugal; Union for a Popular Movement | The Republicans - France; The Republic Onwards! | Renaissance - France; People's Party - Spain; epp; Unity - Latvia; Farmer-Citizen Movement - Netherlands; New Social Contract - Netherlands; Christian Social People's Party - Luxembourg; Estonian Reform Party - Estonia; Civic Democratic Party - Czech Republic; Respect and Freedom Party - Hungary; Pro Patria Union - Estonia; Nationalist Party - Malta; Italy Alive - Italy & 53.2 \\ \cline{2-5}
& \multirow{6}{*}{Rad. Right} & Far-right - Ireland & National Rally - France & 16.2 \\ \cline{3-5}
&  & Far-right - Ireland & Brothers of Italy - Italy & 8.9 \\ \cline{3-5}
&  & Far-right - Ireland & Alternative for Germany - Germany & 7.0 \\ \cline{3-5}
&  & Far-right - Ireland & Voice - Spain & 5.0 \\ \cline{3-5}
& & Alternative for Germany - Germany & Alternative for Germany - Germany & 4.7 \\ \cline{3-5}
 &  & Other & Brothers of Italy - Italy; Party for Freedom - Netherlands; Law and Justice - Poland; Fidesz -- Hungarian Civic Party / Christian Democratic People's Party - Hungary; Enough - Portugal; Danish People's Party - Denmark; National Rally - France; North League - Italy; Sweden Democrats - Sweden; Fidesz -- Hungarian Civic Union - Hungary; Reconquête - France; Voice - Spain; not applicable; Alternative for Germany - Germany; Alliance for the Unity of Romanians - Romania; Finnish Party | True Finns - Finland; Freedom Party of Austria - Austria; Greek Solution - Greece; Sweden Democrats - Sweden ; Latvia First - Latvia; Reconquest - France; Greek Solution – Greece; Victory - Greece; Peoples Association -- Golden Dawn - Greece; Identity-Liberties - France; New Flemish Alliance - Belgium; Flemish Interest - Belgium; Action of Dissatisfied Citizens 2011 - Czech Republic & 58.2 \\ \cline{1-5}
\multirow{3}{*}{Poland} & \multirow{3}{*}{Rad. Left} & The Left (coalition) - Poland & Left Together - Poland & 84.0 \\ \cline{3-5}
&  & Left Together - Poland & Left Together - Poland & 3.8\\ \cline{3-5}
& & Adrian Zandberg & Left Together - Poland & 3.1 \\ \cline{3-5}
\multirow{18}{*}{Poland} & \multirow{3}{*}{Rad. Left} & Maciej Konieczny & Left Together - Poland & 2.3 \\ \cline{3-5}
& & Razem Party - Poland & Left Together - Poland & 2.3 \\ \cline{3-5}
& & Other & Left Together - Poland; Coalition of the Radical Left - Greece & 4.6 \\ \cline{2-5}
& \multirow{6}{*}{Main Left} & The Left (coalition) - Poland & New Left - Poland & 16.7 \\ \cline{3-5}
& & The Left (coalition) - Poland & Democratic Left Alliance - Poland & 16.4 \\ \cline{3-5}
& & The Left (coalition) - Poland & Labour Union - Poland & 16.4 \\ \cline{3-5}
& & The Left (coalition) - Poland & Spring - Poland & 16.4 \\
& & Robert Biedroń & New Left - Poland & 5.5 \\ \cline{3-5}
& & Other & New Left - Poland; Direction -- Social Democracy - Slovakia; Democratic Left Alliance - Poland; Spring - Poland; Social Democrats - Denmark; Social Democratic Party of Germany - Germany; Spanish Socialist Workers Party - Spain; PASOK - Greece; Labour Party - Netherlands; Norwegian Labour Party - Norway; Labour Union - Poland & 28.5 \\ \cline{2-5}
& \multirow{6}{*}{Greens} & Civic Coalition - Poland & The Greens - Poland & 96.1 \\ \cline{3-5}
& & Coalition 15 October - Poland & The Greens - Poland & 1.0 \\ \cline{3-5}
& & Greens - Germany & Alliance 90 / Greens - Germany & 1.0 \\ \cline{3-5}
& & Klaudia Jachira & The Greens - Poland & 1.0 \\ \cline{3-5}
& & Senate Pact Coalition - Poland & The Greens - Poland & 1.0 \\ \cline{2-5}
& Main Right & Donald Tusk & Civic Platform - Poland & 24.9 \\ \cline{3-5}
& \multirow{5}{*}{Main Right} & Third way coalition - Poland & Centre for Poland - Poland & 6.0 \\ \cline{3-5}
& & Civic Coalition - Poland & Civic Platform - Poland & 5.6 \\ \cline{3-5}
& & Civic Coalition - Poland & Modern - Poland & 5.6\\ \cline{3-5}
& & Civic Coalition - Poland & Polish Initiative - Poland & 5.6 \\ \cline{3-5}
\multirow{6}{*}{Poland} & Main Right & Other & Civic Platform - Poland; Poland 2050 - Poland; Polish People's Party - Poland; Christian Democratic Union - Germany; The Republic Onwards! | Renaissance - France; Modern - Poland; Polish Initiative - Poland; Reformist Movement - Belgium; Respect and Freedom Party - Hungary; Party of Liberty and Progress | Flemish Liberals and Democrats - Belgium; Progressive Slovakia - Slovakia; Christian Social Union - Germany; Centre Party - Sweden; Democrats 66 - Netherlands; Fianna Fail - Ireland; Free Democratic Party - Germany; Free Voters - Germany; People's Party for Freedom and Democracy - Netherlands; Union for French Democracy | Democratic Movement - France; Centre for Poland - Poland; Croatian Democratic Union - Croatia; Christian Democratic Union / Christian Social Union - Germany; Austrian People's Party - Austria; Go Italy -- The People of Freedom - Italy; Moderate Party - Sweden; National Coalition Party - Finland; National Liberal Party - Romania; New Democracy - Greece; Slovenian Democratic Party - Slovenia; Social Democratic Party - Portugal; Union for a Popular Movement | The Republicans - France; Farmer-Citizen Movement - Netherlands; New Social Contract - Netherlands; Homeland Union - Lithuania; Danish Social Liberal Party - Denmark; Fine Gael (Family of the Irish) - Ireland; Nationalist Party - Malta & 52.4 \\ \cline{2-5}
& \multirow{6}{*}{Rad. Right} & Law and Justice - Poland & Law and Justice - Poland & 28.5 \\ \cline{3-5}
& & Daniel Obajtek & Law and Justice - Poland & 7.1 \\ \cline{3-5}
& & Jarosław Kaczyński & Law and Justice - Poland & 6.6 \\
& & Mateusz Morawiecki & Law and Justice - Poland & 4.3 \\ \cline{3-5}
& & Confederation Liberty and Independence - Poland & National Movement - Poland & 3.9\\ \cline{3-5}
& & Other & Law and Justice - Poland; Coalition for the Renewal of the Republic -- Liberty and Hope - Poland; United Poland - Poland; Fidesz -- Hungarian Civic Party / Christian Democratic People's Party - Hungary; National Movement - Poland; Alternative for Germany - Germany; National Rally - France; Brothers of Italy - Italy; Voice - Spain; Reconquest - France; Fidesz -- Hungarian Civic Union - Hungary; Party for Freedom - Netherlands; Freedom Party of Austria - Austria; Homeland Union - Croatia; Action of Dissatisfied Citizens 2011 - Czech Republic & 49.6 \\ \cline{1-5}
\multirow{3}{*}{Portugal} & \multirow{3}{*}{Rad. Left} & Bloc of the Left - Portugal & Bloc of the Left - Portugal & 29.0 \\ \cline{3-5}
& & Unified Democratic Coalition - Portugal & Unified Democratic Coalition - Portugal & 23.1 \\ \cline{3-5}
& & Portuguese Communist Party - Portugal & Portuguese Communist Party - Portugal & 13.6 \\ \cline{3-5}
\multirow{20}{*}{Portugal} & Rad. Left & Catarina Martins & Bloc of the Left - Portugal & 10.4 \\ \cline{3-5}
& & João Oliveira & Portuguese Communist Party - Portugal & 6.3 \\ \cline{3-5}
& & Other & Bloc of the Left - Portugal; Portuguese Communist Party - Portugal; Unified Democratic Coalition - Portugal; not applicable; left; Unbowed France - France; Ecology Party -- Greens - Portugal & 17.6 \\ \cline{2-5}
& \multirow{6}{*}{Main Left} & Socialist Party - Portugal & Socialist Party - Portugal & 40.0 \\ \cline{3-5}
& & Marta Temido & Socialist Party - Portugal & 16.9 \\ \cline{3-5}
& & Pedro Nuno Santos & Socialist Party - Portugal & 9.9 \\ \cline{3-5}
& & António Costa & Socialist Party - Portugal & 8.0 \\ \cline{3-5}
& & Pedro Sánchez & Spanish Socialist Workers Party - Spain & 2.9 \\ \cline{3-5}
& & Other & Socialist Party - Portugal; Spanish Socialist Workers Party - Spain; Social Democratic Party of Germany - Germany; Direction -- Social Democracy - Slovakia; Democratic Party - Italy; Social Democrats - Denmark; Luxembourg Socialist Workers' Party - Luxembourg; Socialist Party - France; Social Democratic Party - Romania; Socialist Party [Francophone] - Belgium; Social Democratic Party of Croatia - Croatia; Democratic Left Alliance - Poland; Spring - Poland & 22.2 \\ \cline{2-5}
& \multirow{6}{*}{Greens} & Livre - Portugal & Livre - Portugal & 51.0\\ \cline{3-5}
& & Party for Animals and Nature - Portugal & Party for Animals and Nature - Portugal & 30.0 \\ \cline{3-5}
& & Francisco Paupério & Livre - Portugal & 4.7\\ \cline{3-5}
& & Rui Tavares & Livre - Portugal & 3.7 \\ \cline{3-5}
& & Francisco Guerreiro & Party for Animals and Nature - Portugal & 1.7 \\ \cline{3-5}
& & Other & Party for Animals and Nature - Portugal; Livre - Portugal; Republican Left of Catalonia - Spain; Alliance 90 / Greens - Germany; not applicable; The Greens - Netherlands & 8.9 \\ \cline{2-5}
& \multirow{6}{*}{Main Right} & Democratic Alliance - Portugal & Democratic Alliance - Portugal & 15.0 \\ \cline{3-5}
& & Sebastião Bugalho & Democratic Alliance - Portugal & 12.6 \\ \cline{3-5}
& & Luís Montenegro & Social Democratic Party - Portugal & 9.9 \\ \cline{3-5}
& & Liberal Initiative - Portugal & Liberal Initiative - Portugal & 9.7 \\ \cline{3-5}
& & Ursula von der Leyen & Christian Democratic Union - Germany & 6.6 \\ \cline{3-5}
\multirow{6}{*}{Portugal} & Main Right & Other & Social Democratic Party - Portugal; Liberal Initiative - Portugal; Democratic alliance - Portugal; The Republic Onwards! | Renaissance - France; Democratic and Social Centre -- People's Party - Portugal; Christian Democratic Union - Germany; Nationalist Party - Malta; People's Party - Spain; Reformist Movement - Belgium; Austrian People's Party - Austria; Civic Platform - Poland; Croatian Democratic Union - Croatia; Go Italy -- The People of Freedom - Italy; Moderate Party - Sweden; National Coalition Party - Finland; National Liberal Party - Romania; New Democracy - Greece; Slovenian Democratic Party - Slovenia; Union for a Popular Movement | The Republicans - France; Party of Liberty and Progress | Flemish Liberals and Democrats - Belgium; People's Party for Freedom and Democracy - Netherlands; Fine Gael (Family of the Irish) - Ireland; Democratic Alliance - Portugal; Democratic Convergence | Together for Catalonia - Spain; Progressive Slovakia - Slovakia; Centre Party - Sweden; Democrats 66 - Netherlands; Fianna Fail - Ireland; Free Democratic Party - Germany; Free Voters - Germany; Union for French Democracy | Democratic Movement - France; Popular Monarchist Party - Portugal; epp; Christian Democratic Union / Christian Social Union - Germany; Christian Social Union - Germany; Farmer-Citizen Movement - Netherlands; New Social Contract - Netherlands; Modern - Poland; Poland 2050 - Poland; Polish Initiative - Poland; Polish People's Party - Poland; Estonian Reform Party - Estonia; not applicable; Respect and Freedom Party - Hungary; Democratic Party - Luxembourg & 46.2 \\ \cline{2-5}
& \multirow{5}{*}{Rad. Right} & Enough - Portugal & Enough - Portugal & 33.0 \\ \cline{3-5}
& & André Ventura & Enough - Portugal & 8.2 \\ \cline{3-5}
& & Far-right - Portugal & National Rally - France & 5.8 \\ \cline{3-5}
& & Giorgia Meloni & Brothers of Italy - Italy & 5.2 \\ \cline{3-5}
& & Far-right - Portugal & Enough - Portugal & 4.3 \\ \cline{3-5}
Portugal & Rad. Right & Other & Alternative for Germany - Germany; Fidesz -- Hungarian Civic Party / Christian Democratic People's Party - Hungary; National Rally - France; Enough - Portugal; Voice - Spain; Brothers of Italy - Italy; North League - Italy; Party for Freedom - Netherlands; Law and Justice - Poland; Action of Dissatisfied Citizens 2011 - Czech Republic; Fidesz -- Hungarian Civic Union - Hungary; The Party is Over - Spain; not applicable; Reconquest - France; Alliance for the Unity of Romanians - Romania; Freedom Party of Austria - Austria; Our Homeland Movement - Hungary; Revival - Bulgaria ; Peoples Association -- Golden Dawn - Greece; Identity-Liberties - France & 43.4 \\
\bottomrule
\end{longtable}

\begin{table}[ht!]
\centering
\small

\caption{\textbf{Media outlets in Media Cloud with no election-related news returned. Country corresponds to media outlet's associated country in our analysis.}}
\label{tab:media_outlets_media_cloud_nonews}

\begin{tabular}{
>{\RaggedRight\arraybackslash}p{14cm}
>{\RaggedRight\arraybackslash}p{3cm}
}
\toprule
\textbf{Newspaper} & \textbf{Country}\\
\midrule
exxpress.at, finanzen.at, finanzen.net, fr.de, heute.at, kleinezeitung.at, meinbezirk.at, wko.at & Austria \\ \hline
finanzen.net, fr.de & Germany \\ \hline
rip.ie , thesun.ie & Ireland \\ \hline
niezalezna.pl, o2.pl, pomponik.pl, pudelek.pl, se.pl & Poland \\ \hline
abola.pt, jn.pt, ojogo.pt, zerozero.pt & Portugal \\
\hline
\end{tabular}
\end{table}

\end{document}